\documentclass[
    aps,
    prl,
    twocolumn,
    floatfix,
    superscriptaddress
]{revtex4-1}

\usepackage{subfigure}
\usepackage[utf8]{inputenc}
\usepackage{times}
\usepackage[breaklinks]{hyperref}
\usepackage{color}
\usepackage{siunitx}

\usepackage{amssymb} 
\usepackage{amsmath}
\usepackage{amsthm}
\usepackage{braket}
\usepackage[version=4]{mhchem}

\renewcommand{\vec}[1]{\mathbf{#1}} 

\newcommand{\hatn}{\hat{\mathbf{n}}}

\newcommand{\abs}{\mathrm{\abs}}
\newcommand{\wigner}{\mathcal{W}}

\newcommand{\lpartial}{\overset{\leftarrow}{\partial}}
\newcommand{\rpartial}{\overset{\rightarrow}{\partial}}

\newcommand{\tr}{\mathrm{tr}}

\newcommand{\fraktr}{\mathfrak{Tr}}

\newcommand{\id}{\mathrm{id}}

\newcommand{\dd}{\mathrm{d}}

\newcommand{\Gk}{\underline{G}}

\newcommand{\Gret}{G^{\mathrm{R}}}

\newcommand{\Gadv}{G^{\mathrm{A}}}

\newcommand{\Gles}{G^<}

\newcommand{\bsigma}{\boldsymbol{\sigma}}

\newcommand{\xc}{\Delta_{\mathrm{xc}}}

\newcommand{\soi}{\alpha_{\mathrm{R}}}

\newcommand{\qfield}{\Psi}

\newcommand{\effmass}{m_{\mathrm{eff}}^\ast}
\newcommand{\pgi}{Peter Gr\"unberg Institut and Institute for Advanced Simulation,
Forschungszentrum J\"ulich and JARA, 52425 J\"ulich, Germany}
\newcommand{\aachen}{Department of Physics, RWTH Aachen University, 52056 Aachen, Germany}
\newcommand{\mainz}{Institute of Physics, Johannes Gutenberg University Mainz, 55099 Mainz, Germany}

\newcommand{\vertex}[1]{\underset{#1}{\bullet}}

\usepackage{tikz}
\usetikzlibrary{calc}
\usetikzlibrary{decorations.markings}
\usepackage{xcolor}
\usepackage{graphicx}

\definecolor{jublue}{RGB}{2,61,107}
\definecolor{jublue2}{RGB}{173,189,227}
\definecolor{juhimbeer}{RGB}{235,95,115}
\definecolor{juapricot}{RGB}{250,180,90}

\newcommand{\vertexA}{210}
\newcommand{\vertexB}{90}
\newcommand{\vertexC}{330}

\newcommand{\vertexABr}{120}
\newcommand{\vertexABc}{150}
\newcommand{\vertexABl}{180}

\newcommand{\vertexBCr}{0}
\newcommand{\vertexBCc}{30}
\newcommand{\vertexBCl}{60}

\newcommand{\vertexCAr}{240}
\newcommand{\vertexCAc}{270}
\newcommand{\vertexCAl}{300}

\newcommand{\coords}[1]{
( {cos(#1)} , {sin(#1)} )
}

\newcommand{\oppositeangle}[1]{
{#1 + 180}
}

\newcommand{\drawbasic}{
	\draw[thick] (0,0) circle [radius=1] ;
	\filldraw \coords{\vertexA} circle (2pt) node[align=center, below left] {$\alpha$};
	\filldraw \coords{\vertexC} circle (2pt) node[align=center, below right] {$\gamma$};
	\filldraw \coords{\vertexB} circle (2pt) node[align=center, above] {$\beta$};
}

\newcommand{\firstorderdiagram}{
\raisebox{-1mm}{
\begin{tikzpicture}[decoration={
	markings,
	mark=at position 0.5 with {\arrow{>}}}
] 
    \node (A) at (0,0) {};
	\node (B) at (1,0) {};
	\node (C) at (2,0) {};
	\node (D) at (3,0) {};
	
	\draw[thick, solid, juapricot,postaction={decorate}] (B.center) to[out=70, in=110, distance=8mm] (C.center);
	
	\draw[thick] (A) -- (D);
\end{tikzpicture}
}
}

\newcommand{\amputatedfirstorderdiagram}{
\raisebox{-1mm}{
\begin{tikzpicture}[decoration={
	markings,
	mark=at position 0.5 with {\arrow{>}}}
] 
	\node (B) at (0,0) {};
	\node (C) at (1,0) {};
	
	\draw[thick, solid, juapricot,postaction={decorate}] (B.center) to[out=70, in=110, distance=8mm] (C.center);
	
	\draw[thick] (B.center) -- (C.center);
\end{tikzpicture}
}
}

\newcommand{\moyal}[2]{
	\node (A) at \coords{#1} {};
	\node (B) at \coords{#2} {};
	\draw[thick, solid, juapricot,postaction={decorate}] (A.center) to[out=\oppositeangle{#1}, in=\oppositeangle{#2}, distance=5mm] (B.center);
}

\newcommand{\chediagram}[2]{
\begin{tikzpicture}[decoration={
	markings,
	mark=at position 0.5 with {\arrow{>}}}
] 
	\moyal{#1}{#2}
	\drawbasic
\end{tikzpicture}
}

\newcommand{\kubodiagram}{
\begin{tikzpicture}
    \drawbasic
\end{tikzpicture}
}

\newcommand{\renormalizedkubodiagram}{
\begin{tikzpicture}
\draw[thick] (0,0) circle [radius=1] ;
	\filldraw[juhimbeer] \coords{\vertexA} circle (2pt) node[align=center, below left] {$\bar{\alpha}$};
	\filldraw \coords{\vertexC} circle (2pt) node[align=center, below right] {$\gamma$};
	\filldraw \coords{\vertexB} circle (2pt) node[align=center, above] {$\beta$};
\end{tikzpicture}
}

\newcommand{\smallkubodiagram}{
\raisebox{-.75cm}{\scalebox{0.75}{\kubodiagram}}
}

\newcommand{\smallrenormalizedkubodiagram}{
\raisebox{-.75cm}{\scalebox{0.75}{\renormalizedkubodiagram}}
}

\newcommand{\smallchediagram}[2]{
\raisebox{-.75cm}{\scalebox{0.75}{\chediagram{#1}{#2}}}
}

\newcommand{\dressedA}{
\raisebox{-2.25mm}{
\scalebox{0.75}{
\begin{tikzpicture}[decoration={
	markings,
	mark=at position 0.5 with {\arrow{>}}}
] 
    \node (A) at (0,0) {};
	\node (B) at (.75,0) {};
	\node (C) at (1.5,0) {};
	
	\filldraw (A) circle (2pt) node[align=center, below] {$\alpha$};
	
	\draw[thick, solid, juapricot,postaction={decorate}] (B.center) to[out=70, in=110, distance=6mm] (C.center);
	
	\draw[thick] (A.center) -- (C.center);
\end{tikzpicture}}}
}

\newcommand{\dressedB}{
\raisebox{-2.25mm}{
\scalebox{0.75}{
\begin{tikzpicture}[decoration={
	markings,
	mark=at position 0.5 with {\arrow{>}}}
] 
    \node (A) at (0,0) {};
	\node (B) at (.75,0) {};
	\node (C) at (1.5,0) {};
	
	\filldraw (B) circle (2pt) node[align=center, below] {$\alpha$};
	
	\draw[thick, solid, juapricot,postaction={decorate}] (A.center) to[out=70, in=110, distance=8mm] (C.center);
	
	\draw[thick] (A.center) -- (C.center);
\end{tikzpicture}}}
}

\begin{document}

\setcounter{secnumdepth}{2} 


\title{The chiral Hall effect of magnetic skyrmions from a cyclic cohomology approach}

\author{Fabian R. Lux}

    \email{f.lux@fz-juelich.de}
    \affiliation{\pgi}
    \affiliation{\aachen}
    
\author{Frank Freimuth}

    \affiliation{\pgi}
    
\author{Stefan Bl\"ugel}

    \affiliation{\pgi}
    
\author{Yuriy Mokrousov}

    \affiliation{\pgi}
    \affiliation{\mainz}

\date{\today}

\begin{abstract}
We demonstrate the emergence of an anomalous Hall effect in chiral magnetic textures which is neither proportional to the net magnetization nor to the well-known emergent magnetic field that is responsible for the topological Hall effect. Instead, it appears already at linear order in the gradients of the magnetization texture and exists for one-dimensional magnetic textures such as domain walls and spin spirals. It receives a natural interpretation in the language of Alain Connes' noncommutative geometry. We show that this chiral Hall effect resembles the familiar topological Hall effect in essential properties while its phenomenology is  distinctly different. Our findings make the re-interpretation of experimental data necessary, and offer an exciting twist in engineering the electrical transport through magnetic skyrmions. 
\end{abstract}

\maketitle


Topological magnetic solitons such as magnetic skyrmions represent a class of particle-like magnetization textures which could serve as energy-efficient information bits of the future~\cite{Fert2017,Kang2016}. There are  three important milestones which need to be reached in order to realize this vision, and which are currently an active field of research: the stabilization of room-temperature solitons~\cite{Woo2016, Legrand2019}, their deterministic control~\cite{Litzius2017, Jiang2017} and their deterministic read-out~\cite{Hamamoto2016, Maccariello2018}. With regard to the latter, non-collinear magnetic textures challenge us with broken translational invariance and variations on mesoscopic length scales. This is why the interpretation of experimental transport data is often strongly debated, as has been recently the case for \ce{SrIrO3}/\ce{SrRuO3} bilayers~\cite{Matsuno2016, Ohuchi2018}. While this system may exhibit skyrmionic magnetization textures,  the presence of chiral domain walls and strong spin-orbit coupling (SOC) adds further complexity~\cite{Meng2019} and undermines an {\it a priori} gauge field interpretation of the observed topological Hall effect (THE)~\cite{Bruno2004, Bliokh2005, Franz2014}. 

The THE has been used as a proxy for the detection of a skyrmion phase since the early days of skyrmionics~\cite{Neubauer2009} and various theoretical approaches have been put forward in order to generalize upon its gauge-field interpretation. These are model based extensions of either the gauge-field language in the non-adiabatic regime~\cite{Nakazawa2018,Vistoli2019}, or on a $T$-matrix scattering theory operating in a weak-coupling regime~\cite{Denisov2018,Rozhansky2019}. The only approach which acknowledges the importance of SOC-induced gauge fields to linear order is that by Nakabayashi and Tatara~\cite{Nakabayashi2014}. However, 
previous work on the orbital magnetization of non-collinear spin-textures suggests that any perturbative description in the SOC-strength is eventually insufficient in a regime where no preferred spin reference-frame exists anymore~\cite{Lux2018}. Strong SOC materials such as the above mentioned oxides can fall into this regime as well. Further, it would be desirable to have a formalism which can be generalized to an {\it ab initio} description and is not confined to the model level.

In this work, we close this gap and set the foundation for future progress in the area of electrical transport properties of isolated topological solitons. We do so by rephrasing its constitutive equations in the language of non-commutative geometry~\cite{Onoda2006, Connes1994}. This procedure results in a systematic recipe to incorporate non-collinear magnetism by providing corrections to the conductivity tensor order by order in the gradients of the local magnetization texture. In addition, the transverse components of the conductivity tensor receive a natural interpretation as non-trivial elements in the cyclic cohomology class of the non-commutative phase-space geometry~\cite{Bellissard1994}. We focus on the first non-trivial transverse correction in this expansion which we coin {\it chiral Hall effect} (CHE). We argue that the CHE can provide a non-negligible correction compared to the THE while it can display similar features. 


Central to our approach is the phase-space formulation of quantum mechanics~\cite{Moyal1949, Groenewold1946} and its application to the non-equilibrium Keldysh formalism~\cite{Onoda2006}. In this approach, the non-commutative algebra of quantum operators is translated into the non-commutative $\star$-product
\begin{equation}
    \star \equiv \exp\left\lbrace
    \frac{i \hbar}{2} \left(
    \lpartial_{x^\mu}\rpartial_{p_\mu}
    -
    \lpartial_{p_\mu}\rpartial_{x^\mu}
    \right)
    \right\rbrace,
\end{equation}
acting on phase-space functions. Here, and in the following discussion, we refer to the four-position $(x)^\mu \equiv (c t, \vec{x})$ and momentum $(p)^\mu \equiv (\epsilon / c, \vec{p})$ with respect to the metric signature  $(-+++)$. Both, $x$ and $p$ can now be regarded as classical $c$-numbers. When $H$ denotes the  Hamiltonian and $\underline{\Sigma}$ the self-energy on phase-space, the Dyson equation reads
\begin{equation}
    (\epsilon - H -\underline{\Sigma})\star \underline{G} = \id,
\end{equation}
whose solutions $\underline{G}$ represent the non-equilbrium Keldysh Green's function and encode the physical properties of the system. By the nature of the $\star$-product, a solution of the Dyson equation yields a semiclassical expansion with respect to $\hbar$  with the classical propagator at $\mathcal{O}(\hbar^0)$ represented by
$
    \Gk_0 = (\epsilon - H -\underline{\Sigma})^{-1}
$. In the following, we approximate the self-energy $\underline{\Sigma}$ with a constant broadening $\Gamma$ in the advanced and retarded component $\Sigma^\mathrm{A} = (\Sigma^\mathrm{R})^\ast =i \Gamma$ and the lesser component $\Sigma^< =2 i  f(\epsilon) \Gamma$. 
In the presence of static and homogeneous external electromagnetic fields, the $\star$-product is modified and takes the form~\cite{Onoda2006}
\begin{equation}
\star \equiv \exp\left\lbrace
\frac{i \hbar}{2} \left(
\lpartial_{x^\mu}\rpartial_{p_\mu}
-
\lpartial_{p_\mu}\rpartial_{x^\mu}
+ q F^{\mu\nu} \lpartial_{p^\mu}\rpartial_{p^\nu}
\right)
\right\rbrace.
\end{equation}
This equation introduces the electromagnetic field tensor $F_{\mu\nu}$ which follows the usual conventions. In particular, $c F_{0i} = E_i $, where $c$ is the speed of light and $q=-e$ denotes the electric charge.
 Within the semiclassical formalism, the electric four-current density $j^\mu \equiv ( c \rho , \vec{j})$ is given by
\begin{equation}
j^\mu \equiv \frac{e}{2} \Im\int \frac{\dd p}{(2 \pi)^{d+1}\hbar^d} ~ \tr
\left\lbrace  \frac{\partial}{\partial p_\mu} (\Gret)^{-1\star} 
~\overset{\star}{,}~
\Gles \right\rbrace,
\end{equation}
where $\lbrace  
~\overset{\star}{,}~
\rbrace$ indicates the anti-commutator with respect to the $\star$-product and which fulfills the continuity equation $\partial_\mu j^\mu = 0 $ (see the Supplemental Material). We define the net current density as the average
$
    \braket{\vec{j}} \equiv V^{-1} T^{-1} \int \dd x ~\vec{j} (\vec{x}),
$
where $T$ represents the time-interval of measurement and $V$ the $d$-dimensional volume of the sample. The conductivity tensor is then defined as the derivative
$
    \sigma_{kl} = \left.  \partial \Braket{j}_k / \partial E_l \right|_{\vec{E}\to 0}
$.
In order to determine this derivative, we expand $\Gk \to \Gk + \Gk_{\vec{E}}$ to find the first-order correction due to the electric field (see Supplemental Material):
\begin{equation}
\Gk_{\vec{E}} = \frac{i \hbar q}{2} \Big(
\Gk \star \nabla_\vec{p} \Gk_0^{-1} \star \partial_\epsilon \Gk
-
\Gk \star \partial_\epsilon \Gk_0^{-1} \star  \nabla_\vec{p} \Gk
\Big) \cdot \vec{E},
\end{equation}
where the $\star$-product and the Green's function is now evaluated at zero-field. The conductivity can then be split into two contributions $\sigma_{kl} = \sigma^\mathrm{sea}_{kl}  + \sigma^\mathrm{surf}_{kl} $. Employing the convention
\begin{equation}
    \fraktr \lbrace \bullet \rbrace \equiv   
    \int \frac{ \dd p}{(2 \pi )^{d+1}\hbar^d}~\tr \lbrace \bullet \rbrace,
\end{equation}
we arrive at the deformed Kubo-Bastin formula
\begin{align}
    \sigma^\mathrm{sea}_{kl} &= \hbar e^2\Re ~
    \fraktr_\mathrm{sea} \Braket{  \Gret \star v_k \star \Gret  \star v_l \star \Gret - ( k \! \leftrightarrow \! l) },
    \\
     \sigma^\mathrm{surf}_{kl} &= \hbar e^2  \Re ~
    \fraktr_\mathrm{surf} \Braket{
    v_k \star ( \Gadv - \Gret ) \star v_l \star \Gadv 
    }.
    \notag \\
\end{align}
Here, we denote $\fraktr_\mathrm{sea} \lbrace \bullet \rbrace = \fraktr
\lbrace f(\epsilon)~ \bullet \rbrace
$ and $\fraktr_\mathrm{surf} \lbrace \bullet \rbrace = \fraktr
\lbrace f'(\epsilon) ~\bullet \rbrace
$, where $f$ is the Fermi distribution.
The derivation of these equations heavily relies on certain algebraic properties of the $\star$-product. In particular its associativity and the cyclic property of the phase-space, $
     \fraktr \Braket{  A \star B}  =  \fraktr \Braket{  B \star A} 
$,
which would not be valid without the real-space averaging. 

\begin{figure}[t]
    \centering
    \includegraphics[width=\linewidth]{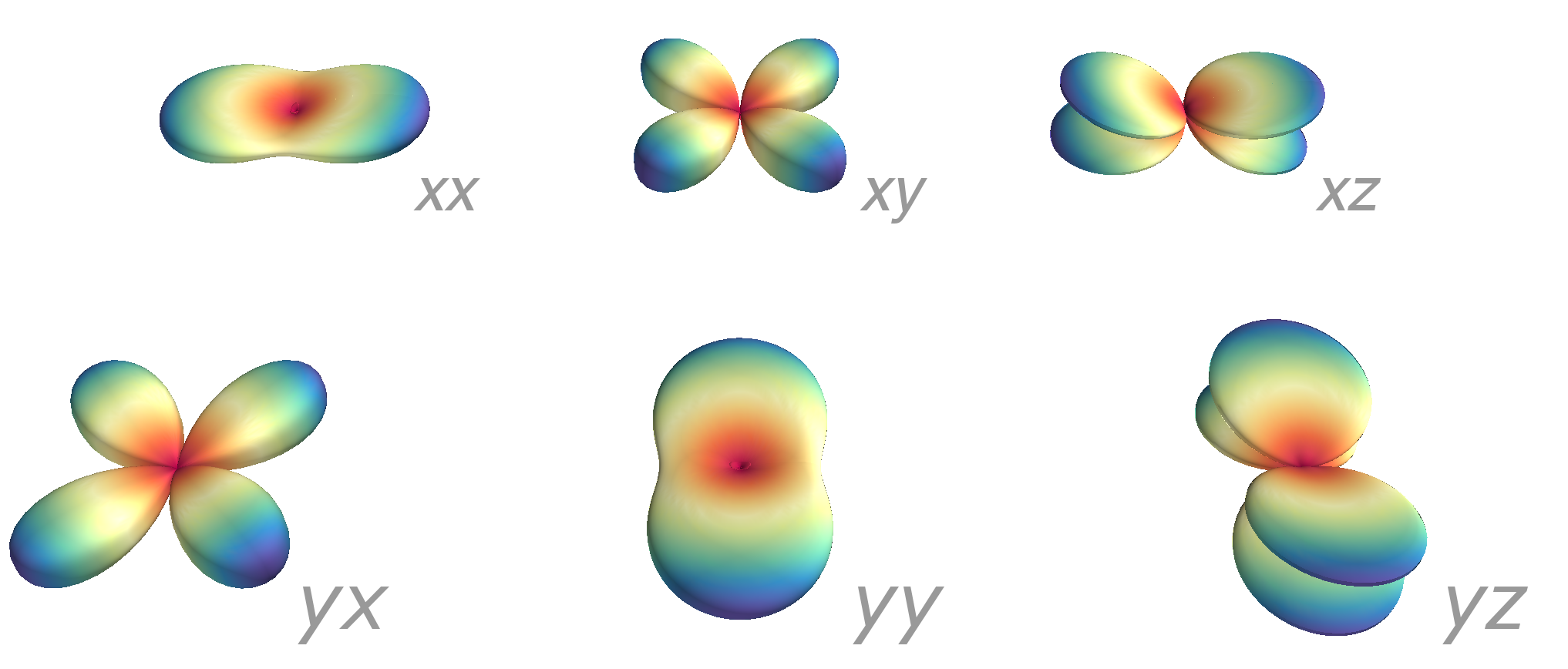}
    \caption{Angular dependence of the chiral Hall effect in the Rashba model. The shown matrix  represents the normalized angular magnetization dependence of the components of the CHE which couple to the real-space derivatives $\partial_i \hat{n}_j$, where $i \in \lbrace x,y \rbrace$ is the row-index and $j\in \lbrace x,y,z \rbrace$  is the column index of the figure ($\hbar\soi = \SI{1.41}{\electronvolt\angstrom}$, $\xc = \SI{0.5}{\electronvolt }$, $\mu = 0$, $\Gamma= \SI{50}{\milli\electronvolt }$, $\effmass =  3.81~ m_\mathrm{e}$). All shapes have an inversion center resulting from Onsager reciprocity. }
    \label{fig:angular_dependence}
\end{figure}

We now want to comment on the geometrical interpretation of the anomalous Hall effect for the special case $d=2$ of two space dimensions. For this, we introduce the Maurer-Cartan like objects
$
    \theta_\mu \equiv \left(\Gk \star \frac{\partial}{\partial p^\mu} (\Gk)^{-1\star } \right)
$, and by defining 
\begin{equation}
    \varphi^{2}(\theta_0, \theta_1, \theta_2 ) 
    \equiv  \Re ~\fraktr_\mathrm{sea} \Braket{  \epsilon^{\mu_0 \mu_1 \mu_2} \theta_{\mu_0}^\mathrm{R} \star \theta_{\mu_1}^\mathrm{R}  \star \theta_{\mu_2}^\mathrm{R} },
\end{equation}
where $\mu \in \lbrace \epsilon, p_x, p_y \rbrace$, we can write the Hall conductivity as
\begin{equation}
     \sigma^\mathrm{sea}_{xy} = 
     \frac{\hbar e^2}{3} \Re ~
    \varphi^2(\theta_0, \theta_1, \theta_2)
    .
\end{equation}
As shown explicitly in the Supplemental Material, $\varphi^2$ represents a cyclic cocycle, i.e., it represents the cyclic cohomology class of the electronic system (see ~\cite{Khalkhali2004} for an introduction to the subject). It can be seen as a generalization of the Thouless-Kohmoto-Nightingale-Nijs (TKNN) formula. In fact, for insulators in the semiclassical limit $\hbar\to0$, one retains the form of the generalized TKNN formula
$\sigma^\mathrm{sea}_{xy} = \frac{e^2}{h} N_2$~\cite{Thouless1982, Qi2011}, where $N_2$
is the Chern character of the electronic system. Also in the general case, a smooth variation of the Green's function $\delta \Gret$ will not alter $\varphi^2$ for insulators at zero temperature, i.e., $\delta \varphi^2 = 0$. A proof of this statement can be obtained by  generalizing an argument of~\cite{Qi2008} (see Supplemental Material). That it should be quantized can be understood from the considerations in non-commutative Chern-Simons theory~\cite{Sheikh-Jabbari2001}. Symbolically, we will write the Kubo 2-cocycle as
\begin{equation}
    \varphi^2(\theta_0, \theta_1, \theta_2) =\Re ~\fraktr_\mathrm{sea}~ \langle \epsilon^{\alpha\beta\gamma} \smallkubodiagram  \rangle,
    \label{eq:cyclic_kubo}
\end{equation}
which puts an emphasis on the cyclic invariance of the trace operation and on its geometrical origin. Here, the solid lines represent retarded Green's functions while the vertices represent derivatives of the inverse propagator, i.e.,
$
    \underset{\alpha}{\bullet} \equiv \partial_{p^\alpha} (\Gret_0)^{-1}
$.
In order to translate the diagram into an equation it is to be read in clockwise direction. The next-order correction (the ``one-loop'' level) can be obtained by expanding the Green's functions to first order in the gradients~\cite{Onoda2006}
$
    \Gret = \Gret_0 + \frac{i \hbar }{2} \Pi^{ij}
    \Gret_0 \partial_i (\Gret_0)^{-1} \Gret_0 \partial_j (\Gret_0)^{-1}\Gret_0 ,
$
where the tensor $\Pi$ encodes the first-order expansion of the $\star$-product. For time-independent Hamiltonians, the indices $i$ and $j$ therefore run over the components of $\vec{x}$ and $\vec{p}$. Diagrammatically, this correction can be expressed as
\begin{equation}
\Gret =   \Gret_0 +  \firstorderdiagram . 
\end{equation}
Inserting this first-order expansion into Eq.~\ref{eq:cyclic_kubo}, expanding the $\star$-products and retaining only first-order contributions in $\hbar$ leads to the full set of diagrams presented in the Supplemental Material. All corrections can be incorporated into a single renormalized (dressed) four-momentum vertex $\bar{\alpha}$, concisely written as
\begin{align}
     \underset{\bar{\alpha}}{\bullet} =& ~ 3~ \Big( \! \! \dressedA  -
      \partial_{p_\alpha}
      \scalebox{.75}{\amputatedfirstorderdiagram }\Big) -2\dressedB.
\end{align}
Going from the zeroth order to the one-loop corrections then amounts to the replacement $\underset{\alpha}{\bullet} \to \underset{\bar{\alpha}}{\bullet} $ in Eq.~\ref{eq:cyclic_kubo}.
Since the renormalized vertex is by construction linear in the gradients of the magnetization texture, we refer to the resulting Hall signal $\sigma^\mathrm{che}_{xy}$ as the {\it chiral Hall effect}. 


\begin{figure}[t]
    \centering
    \includegraphics[width=\linewidth]{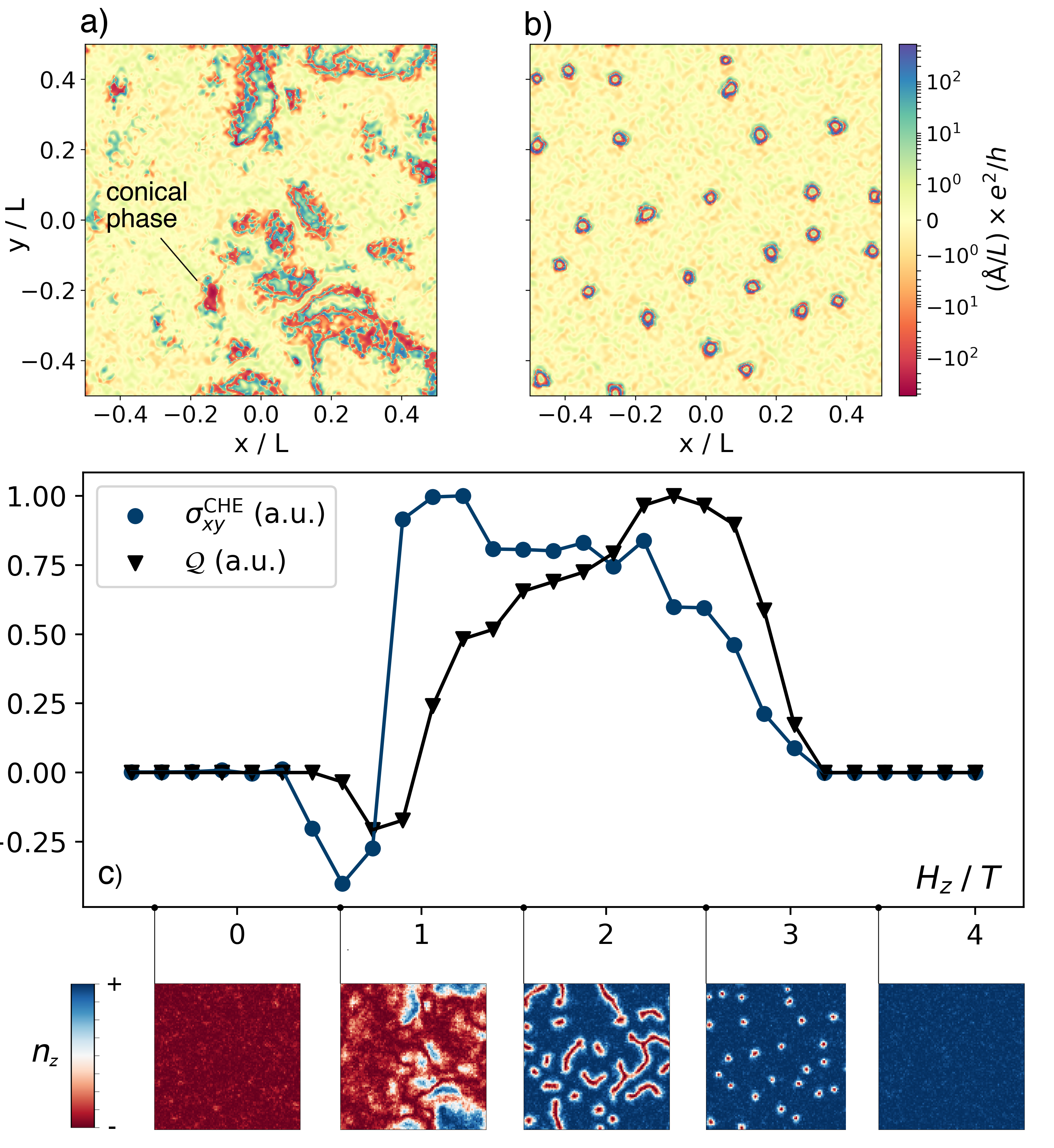}
    \caption{Chiral Hall effect in a chiral Heisenberg magnet. A 2D-system of side-length $L$ containing $100\times100$ spins is simulated at $T=\SI{1}{\kelvin}$ using the {\it Spirit} code. Starting from the polarized state at  $H_z=\SI{-4}{\tesla}$, the magnetic field is increased step-by-step. a) At $H_z=\SI{0.57}{\tesla}$ the magnetization starts to switch. Shown is the density of the CHE which displays strong resonances at local conical phases which have no topological charge. b) At $H_z=\SI{2.53}{\tesla}$, the system is dominated by magnetic skyrmions which exhibit a ring-like CHE signature. c) Tracing the evolution of the average CHE and the topological charge $\mathcal{Q}$ one finds a correlated signal (the corresponding magnetic state is shown in the inset below). Yet, by its nature, the CHE is not caused by $\mathcal{Q}$. }  
    \label{fig:hysteresis}
\end{figure}

Consider a generic time-independent two-band Hamiltonian of the form
$
    H \equiv d_\mu \sigma^\mu
$,
where $\mu \in \lbrace 0,1,2,3 \rbrace$,  $\sigma^\mu =(\id, \bsigma)$ and $d^\mu = (d^0, \vec{d})$ in the $(+---)$ metric convention. We consider $H$ as the Wigner transformation of some quantum Hamiltonian and therefore, $d^\mu$ is a function of the phase space coordinates $\vec{x}$ and $\vec{p}$. Defining the four-vector
$
    g^\mu = (g^0, \vec{g}) \equiv (\epsilon - d^0 + i \Gamma, \vec{d} ) ,
$
the retarded Green's function for this Hamiltonian is given by $\Gret_0 =  g_\mu \sigma^\mu /g^\nu g_\nu$.
With these conventions, the first order gradient correction can be obtained via a suitable renormalization of the current vertex as described above. The renormalized diagram evaluates to
\begin{align}
   \tr~\epsilon^{\alpha\beta\gamma} \smallrenormalizedkubodiagram =& \frac{-4i}{(g^\mu g_\mu )^2} ~
  \Big\lbrace 
  \vec{d}\cdot
    \left( \partial_{p_x} \vec{d} \times  \bar{\vec{v}}_{p_y} \right) 
\notag \\
& + \vec{d}\cdot
    \left( \bar{\vec{v}}_{p_x} \times \partial_{p_y}  \vec{d}  \right)
\notag \\
& + \bar{v}^0_\epsilon ~
\vec{d} \cdot 
\left( \partial_{p_x} \vec{d} \times \partial_{p_y} \vec{d} \right) 
\notag \\
& + g^0 ~ 
\bar{\vec{v}}_\epsilon \cdot 
\left( \partial_{p_y} \vec{d} \times \partial_{p_x} \vec{d} \right) \Big\rbrace ,
\end{align}
where we have defined $
    \bar{v}^\mu_\alpha \equiv 
    \frac{1}{2} \tr~ \underset{\alpha}{\bullet}~ \sigma^\mu  
$ . 
Evaluating the trace for the coupling to $\bar{v}^0_\epsilon$ yields
\begin{equation}
     \bar{v}^0_\epsilon =  -\frac{1}{2} \partial_\epsilon  \tr \Big\lbrace \! \!
    \amputatedfirstorderdiagram
    \!\! \Big\rbrace =
      \partial_\epsilon\frac{2|\vec{d}|^3}{g_\mu g^\mu}
    \sum_i \Omega^{x_i p_i},
\end{equation}
where we now have introduced the mixed-space Berry  curvature of a two-level system, which can be written as $\Omega^{ij} = - \vec{d}\cdot(\partial_i \vec{d}\times \partial_j\vec{d}) / (2 |\vec{d}|^3 )$.  The term proportional to $\bar{v}^0_\epsilon$ thus describes a coupling of the momentum space Berry curvature $\Omega^{p_x p_y} $ to the mixed space Berry curvatures $\Omega^{x_i p_i}$ $-$ a connection which was speculated on before (but not shown to exist) for the chiral orbital magnetization of isolated skyrmions~\cite{Lux2018}. The appearance of a term like this is remarkable, since the semiclassical dynamics of insulators would be well captured by the St\v{r}eda formula with a modified density of states given by the Pfaffian $\mathcal{D} =  \left|\mathrm{Pf} (\boldsymbol{\Omega}- \boldsymbol{\Pi}) \right|/{(2\pi)^2}$~\cite{Streda1982,Xiao2005,Xiao2010} and  a coupling like $\Omega^{p_x p_y}\Omega^{x_i p_i}$ is not present in the two-dimensional expansion of the Pfaffian.

Finally, we demonstrate the existence of the CHE by performing an explicit numerical simulation for a spin-polarized electronic system with spin-orbit interaction.
Namely, we focus our analysis on the two-dimensional magnetic Rashba model, which is well-suited to describe the effect of SOC in interfacial systems: 
\begin{equation}\label{rashba:ham}
H = \frac{\vec{p}^2}{2 \effmass} + \soi ( \boldsymbol{\sigma} \times \vec{p}  )_z + \xc\ \boldsymbol{\sigma} \cdot \hat{\mathbf{n}}(\vec{x}),
\end{equation}
where $\effmass$ is the electron's (effective) mass, $\boldsymbol{\sigma}$ denotes the vector of Pauli matrices, $\soi$ is the Rashba spin-orbit coupling constant, and $\xc$ is the strength of the local exchange field. In general, the CHE can be formulated in a tensorial way as
\begin{align}
    \sigma^\mathrm{che}_{xy}[\hatn]
    = \frac{1}{V} \int \dd \vec{x}~ (\sigma^\mathrm{che}_{xy})^{ij}~(\hatn)  \partial_i \hat{n}_j,
\end{align}
where $(\sigma^\mathrm{che}_{xy})^{ij}(\hatn)$ depends only on the direction of $\hatn$, but not on its derivatives.  In Fig.~\ref{fig:angular_dependence}, we present the zero-temperature directional dependence of this tensor for the case of $\hbar\soi = \SI{1.41}{\electronvolt\angstrom}$, $\xc = \SI{0.5}{\electronvolt }$, $\Gamma= \SI{50}{\milli\electronvolt }$ and $\effmass = 3.81~ m_\mathrm{e}$. The chemical potential was set to $\mu = \SI{0}{\electronvolt}$. All shapes have an inversion center due to the local Onsager reciprocity $(\sigma^\mathrm{che}_{xy})^{ij}~(-\hatn) =(\sigma^\mathrm{che}_{xy})^{ij}~(\hatn) $, proved in the Supplemental Material (generalizing results from~\cite{Banyai1994}). Notably, the angular dependence does not resemble a dominant $n_z^2$ behavior in $(\sigma^\mathrm{che}_{xy})^{xx}(\hatn)$ and $(\sigma^\mathrm{che}_{xy})^{yy}(\hatn)$ as would be expected for the Berry phase term which was alluded to above. This means that the Berry curvature contribution to the CHE is subdominant in this case, and it should therefore be rather interpreted as a strongly non-adiabatic effect. 

It lies in the power of our approach that once the directional dependence of the CHE tensor is known, it can be integrated for arbitrary magnetization textures. We chose to perform a numerical experiment using the spin dynamics code {\it Spirit}~\cite{Mueller2019}, and perform a hysteresis loop experiment for a Heisenberg magnet with ferromagnetic nearest neighbor exchange $J=\SI{1}{\milli\electronvolt }$, Dzyaloshinskii-Moriya interaction $D=\SI{0.5}{\milli\electronvolt }$ and spin moment  $\mu_s=2\mu_B$ at a temperature of $T= \SI{1}{\kelvin}$. This system stabilizes skyrmions in between $\SI{2}{\tesla}$ and $\SI{3}{\tesla}$ (for further details we refer to the Supplemental Material). This kind of experiment is commonly conducted in materials where skyrmions are so small that optical techniques are not available for their detection. Instead, the presence of skyrmions is inferred from an additional feature in the anomalous Hall signal which is commonly attributed to the topological Hall effect.

The results of this numerical experiment, presented in Fig.~\ref{fig:hysteresis}(a), demonstrate the emergence of the chiral Hall effect in the switching process and its correlation  with the build-up of the topological charge. However, unlike the topological Hall effect it is not caused by this charge. This becomes evident when we look closer at the field strength of $H = \SI{0.57}{\tesla}$, at which local conical phases appear in the sample. Remarkably, the CHE shows a very large magnitude in the areas which exhibit conical spin spirals (see the inset of Fig.\ref{fig:hysteresis}(b) of the main text, and Fig.~(3) in the Supplemental Material), while the topological charge is still zero in this case. Within the field regime where magnetic skyrmions are stabilized, the CHE shows a non-trivial behavior, and manifests itself in ring-like textures in the local current density, see the inset of Fig.\ref{fig:hysteresis}(c).

Overall, our findings call for a re-interpretation of common transport experiments in chiral magnets. In the past, these mostly relied on phenomenological arguments based on the adiabatic theory of the topological Hall effect, which is already known to be insufficient in many cases \cite{Nakazawa2018,Vistoli2019,Denisov2018,Rozhansky2019}. Our formalism represents the first systematic derivation of corrections to the anomalous Hall effect of collinear magnets, which are in principle not restricted to the model level or the weak-coupling regime, and which can be certainly reformulated in the language of {\it ab initio}  electonic structure. 

We believe that the CHE will provide an essential degree of freedom in engineering the electrical transport through magnetic skyrmions and non-collinear magnetic textures in general. This is  emphasized by the fact that according to our analysis the non-vanishing CHE emerges already for one-dimensional spin textures such as spin-spirals and domain walls~\cite{Lux2019}. In materials with pronounced spin-orbit interaction the CHE  can be a significant contribution to the overall Hall signal. This might be the case for example in the aforementioned \ce{SrIrO3}/\ce{SrRuO3} bilayers or the celebrated B20-compound \ce{MnSi}, where scattering off the cone-phase has been speculated to lead to a Hall effect  before~\cite{Meynell2014}. But also in antiferromagnetic systems of~e.g.~the \ce{Mn3X}-family, which exhibit the ``topological" Hall effect, an investigation of the CHE can be of great interest~\cite{Kuebler2014, Nayak2016}.

Importantly, we have embedded the approach to the CHE into the rich context of non-commutative geometry which lends itself to a deep interpretation. This connection is fruitful for two reasons. Firstly, its language might shed some light on how super-lattices of magnetic skyrmions can induce topological states in magnetic insulators as they change the topology of the underlying non-commutative phase space. Secondly, electronic transport in non-collinear magnets on its own represents an experimentally well-established field, where theoretical ideas from non-commutative geometry and non-commutative Chern-Simons theory could be put under scrutiny, while they remain rather elusive in the context of high-energy physics where they have been studied intensively 20 years ago~\cite{Seiberg1999, Douglas2001, Szabo2003}.

\begin{acknowledgements}
We thank Vincent Cros for fruitful discussions and  Gideon P. Müller for useful advice regarding the {\it Spirit} code. Further, we  acknowledge  funding  under SPP 2137 ``Skyrmionics" (project  MO  1731/7-1)  of  Deutsche  Forschungsgemeinschaft (DFG) and also gratefully acknowledge the J\"ulich Supercomputing Centre and RWTH Aachen University for providing computational resources under project jiff40.
\end{acknowledgements}


\hbadness=99999 
\bibliographystyle{apsrev4-1}

%

\cleardoublepage
\appendix

\onecolumngrid

\begin{center}
    {\bf \large Supplemental Material: The chiral Hall effect of magnetic skyrmions from a general cyclic cohomology approach}
\end{center}

\section{Derivation of the deformed Kubo formula and the chiral Hall effect}

\subsection{Definition of the Wigner transformation}

We follow the conventions in~\cite{Onoda2006}. Given a function $o(x_1, x_2)$ with a dependence on two space-time points $x_1$ and $x_2$ one constructs the auxiliary function $o_x (y)$, where $x = (x_1 + x_2) / 2$ and $y = x_1 - x_2$. The Wigner transformation $\wigner$ of $o$ is then defined as the Fourier transformation
\begin{equation}
    \wigner [ o] (x,p) = \int \dd^{d+1} y ~ o_x(y) ~e^{-i y_\mu p^\mu / \hbar} . 
\end{equation}
As the Fourier transformation is invertible, one can express  $o(x_1, x_2)$ in terms of its Wigner transformation as
\begin{equation}
    o_x(y) = \frac{1}{(2\pi \hbar)^{d+1}}\int \dd^{d+1} p ~  \wigner [ o] (x,p)  ~e^{+i y_\mu p^\mu / \hbar}.
\end{equation}
For a space-time convolution of the form 
$
    (o_1 \ast o_2 ) (x_1, x_2 ) \equiv \int \dd y ~o(x_1, y) o(y, x_2)
$,
the Wigner transformation is given by the $\star$-product expression
\begin{equation}
    \wigner[ o_1 \ast o_2 ] (x,p) = \wigner[o_1](x,p) \star \wigner[o_2](x,p)
\end{equation}
with the $\star$-product defined as in the manuscript.

\subsubsection{Expectation values of operators}

A time-averaged single particle operator $\hat{A}^{(1)}$ in second quantized form is given by
\begin{equation}
	\hat{A} =  \frac{1}{V T}\sum_{\sigma\sigma'}\int \dd t\int \dd^d r \int \dd^d r '~
	\qfield_\sigma^\dagger (\vec{r},t)
	\Braket{
	\vec{r}\sigma
	|
		\hat{A}^{(1)}
	|
	\vec{r}' \sigma'
	}
	\qfield_{\sigma'}(\vec{r}',t) ,
\end{equation}
where
$
\qfield_\sigma^\dagger (\vec{r},t)=  \sum_{\vec{k}}  e^{-i \vec{k}\cdot \vec{r}}a_{\vec{k},\sigma}^\dagger (t) 
$.
We now define the integration kernel 
$
A_{\sigma\sigma'}(\vec{r}t, \vec{r}'t') = \delta(t-t')
		\Braket{
		\vec{r}\sigma
		|
		A^{(1)}
		|
			\vec{r}' \sigma'
	}
$. Therefore
\begin{align}
    \braket{\hat{A}}_q =-i\hbar \frac{1}{VT}
    \int \dd^{d+1} x_1 \lim_{x_2 \to x_1} ~ \tr ~ (A \ast \Gles) (x_1, x_2) \equiv -i\hbar
   \Braket{ \lim_{x_2 \to x_1} ~  \tr ~ (A \ast \Gles) (x_1, x_2) }
\end{align}
where we have identified the lesser Green's function $\Gles_{\sigma \sigma'} (x_1, x_2) = i \Braket{\qfield^\dagger_{\sigma'}(x_2)  \qfield_{\sigma}(x_1)}_q / \hbar~$ and where the index $q$ indicates the quantum statistical expectation value. The bracket on the r.h.s. again denotes the space-time averaging as in introduced in the manuscript. Based on the inverse Wigner-transformation, one finds
\begin{align}
    \lim_{x_2 \to x} ~  \tr ~ (A \ast \Gles) (x, x_2)  &=
    \lim_{y \to 0}
    \frac{1}{(2\pi \hbar)^{d+1}}\int \dd^{d+1} p ~  \tr~\wigner [ A \ast \Gles] ( x ,p)  ~e^{+i y_\mu p^\mu / \hbar}
    \notag \\
    & = 
    \frac{1}{(2\pi \hbar)^{d+1}}\int \dd^{d+1} p ~  \tr~\lbrace\wigner [ A ]( x ,p) \star \wigner[ \Gles] ( x ,p) \rbrace .
\end{align}
Usually, we neglect the explicit notation of $\wigner$ and will do so again in the following. In summary, the time-averaged expectation value of the single particle operator $\hat{A}$ is given by
\begin{equation}
    \braket{\hat{A}}_q =- i \fraktr~ \braket{A \star \Gles},
\end{equation}
with $A = A(x,p)$ and $\Gles = \Gles (x,p)$. Using the cyclic property of the phase-space trace, it would be indeed possible to drop the $\star$-product in this expression. It his however often beneficial to keep it until the point where semiclassical approximations are adopted. For Hermitian operators, i.e., for $\hat{A}^\dagger = \hat{A}$, and using the property that $(\Gles)^\dagger = - \Gles$, one finds
\begin{equation}
    \braket{\hat{A}}_q = \Im ~\fraktr~ \braket{A \star \Gles}.
\end{equation}

\subsection{Useful relations for Wigner-space Green's functions}

\subsubsection{Analytic properties of the Green's function}
We consider the left and right Dyson equations
\begin{align}
 \underline{G}_\mathrm{l} \star (\epsilon - H -\underline{\Sigma}) &= \id
 \\
    (\epsilon - H -\underline{\Sigma})\star \underline{G}_\mathrm{r} &= \id .
\end{align}
Since the $\star$-product is associative, we find that $\underline{G}_\mathrm{l} = \underline{G}_\mathrm{r} = \underline{G}$. This can be seen as follows:
\begin{align}
      \underline{G}_\mathrm{r} = \id \star \underline{G}_\mathrm{r} = (\underline{G}_\mathrm{l} \star (\epsilon - H -\underline{\Sigma})) \star \underline{G}_\mathrm{r} 
     = \underline{G}_\mathrm{l} \star ( (\epsilon - H -\underline{\Sigma}) \star \underline{G}_\mathrm{r}  )
     =  \underline{G}_\mathrm{l} \star \id  
     =  \underline{G}_\mathrm{l} .
\end{align}
Further, if the self-energy fulfills $\Sigma^\mathrm{R} = (\Sigma^\mathrm{A})^\dagger$ and  $\Sigma^< = -(\Sigma^<)^\dagger$, then the Green's function obeys
\begin{align}
    \Gret = (\Gadv)^\dagger ,
    \hspace{5mm}
    \Gles = - (\Gles)^\dagger .
\end{align}
At zeroth order in $\hbar$, the Dyson equation reduces to 
$
    (\epsilon - H -\underline{\Sigma}) \Gk_0 = \id 
$.

\subsubsection{The Green's function under time-reversal}

We consider a Hamiltonian of the form 
\begin{align}
    H [ \hatn ] \equiv H_0 + \xc \hatn\cdot\bsigma,
\end{align}
where $H_0$ is invariant under the time reversal operation $\mathrm{T}$. On phase-space, we decompose  $\mathrm{T} = \mathcal{U} \star K $ into a $\star$-unitary part $\mathcal{U}$ and the complex conjugation $K$, and define its action on phase-space functions by
\begin{equation}
    \mathrm{T} \star \mathcal{O}(\vec{x},t; \vec{p}, \epsilon ) \star \mathrm{T}^{-1} \equiv  \mathcal{U} \star \mathcal{O}(\vec{x},t; \vec{p}, \epsilon )^\ast \star \mathcal{U}^\dagger
    \equiv  u  \mathcal{O}(\vec{x},-t; -\vec{p}, \epsilon )^\ast  u^\dagger,
\end{equation}
where the unitary matrix $u$ acts on the spin degree of freedom. Here, we require that
$
    \mathrm{T} \star \bsigma \star \mathrm{T}  = - \bsigma
$.
The Hamiltonian therefore transforms as $ H [ \vec{p}, \hatn ]  = \mathrm{T}\star H [ -\vec{p}, - \hatn ]  \star\mathrm{T}^{-1}$. It then follows directly from the Dyson equation, that
\begin{align}
    \Gret_{0,\hatn } (\vec{x},t; \vec{p}, \epsilon ) &= + \mathrm{T} \star \Gadv_{0,-\hatn }(\vec{x},-t; -\vec{p}, \epsilon ) \star \mathrm{T}^{-1}, \\ 
    \Gadv_{0,\hatn } (\vec{x},t; \vec{p}, \epsilon ) &= + \mathrm{T}\star \Gret_{0,-\hatn }(\vec{x},-t; -\vec{p}, \epsilon ) \star \mathrm{T}^{-1}, \\ 
    \Gles_{0,\hatn }(\vec{x},t; \vec{p}, \epsilon ) &= -           \mathrm{T}\star \Gles_{0,-\hatn }(\vec{x},-t; -\vec{p}, \epsilon ) \star \mathrm{T}^{-1} .
 \end{align}
In the absence of electromagnetic fields, the $\star$-product is invariant under time-reversal, and these relations generalize to the full propagator $\Gk$. For $A = A(x,p)$, it then follows that
\begin{align}
      \Re~ \tr~ \lbrace A \star \Gles_{\hatn}(\vec{x},t ; \epsilon, \vec{p}) \rbrace&= - \Re~ \tr~ \lbrace
    ( \mathcal{U}^\dagger \star A\star \mathcal{U})^\ast \star \Gles_{-\hatn}(\vec{x},-t ; \epsilon, -\vec{p})
    \rbrace , \\
     \Im~ \tr~ \lbrace A \star \Gles_{\hatn}(\vec{x},t ; \epsilon, \vec{p}) \rbrace&= + \Im~ \tr~ \lbrace
    ( \mathcal{U}^\dagger \star A\star \mathcal{U})^\ast \star \Gles_{-\hatn}(\vec{x},-t ; \epsilon, -\vec{p}),
    \rbrace .
\end{align}
This is a generalization of the result in~\cite{Banyai1994}.
In particular, this implies that the expectation value of Hermitian operators (such as the density of states $\sim \Im~\tr~\Gles$) is invariant under time-reversal as long as the $\star$-product remains $\mathrm{T}$-invariant.

\subsection{Current operator and continuity equation}

Following the relativistic notation, we define the Hermitian four-velocity operator as
\begin{equation}
    v^\mu \equiv -\frac{1}{2} ~ \big( \partial_{p_\mu}\Gk^{-1\star}  
    + \mathrm{h.c.} \big) .
\end{equation}
The electric four-current density $j^\mu=(c\rho, \vec{j})$ can then be obtained from the phase-space expectation value of $v_\mu$ using its local expression, i.e.,
\begin{equation}
j^\mu \equiv \frac{e}{2} \Im~\fraktr~
\left\lbrace  \frac{\partial}{\partial p_\mu} (\Gret)^{-1\star} 
~\overset{\star}{,}~
\Gles \right\rbrace,
\end{equation}
where $e$ is the elementary charge. From the Dyson equation one obtains
\begin{align}
    \partial_{x^0} j^0 &= -e \Im ~ \fraktr ~ \frac{1}{i \hbar} [ H \overset{\star}{,} \Gles]
    \\
    \mathrm{div}~\vec{j}&= +e \Im ~ \fraktr ~ \frac{1}{i } [ H \overset{(\partial_\hbar \star)}{,} \Gles].
\end{align}
As can be proven order by order in $\hbar$, we have the identity
$
    \fraktr [ H \overset{\star}{,} \Gles] / ( i\hbar)
    =\fraktr [ H \overset{(\partial_\hbar \star)}{,} \Gles] / i
$ for Hamiltonians of the form $H = p^2 / (2m) + V(x)$
and therefore, the continuity equation
$\partial_{x^\mu} j^\mu = 0$ is fulfilled. That a continuity equation like this should be fulfilled was claimed in~\cite{Onoda2006} but not proved.

\subsection{Perturbation of the Green's function}

Applying a partial derivative with the property $\partial \star = 0$ to the Dyson equation and using the associativity of the $\star$-product, one finds the matrix identity
\begin{equation}
    \partial \Gk = - \Gk \star \partial \Gk^{-1\star }  \star \Gk = - \Gk \star \partial \Gk_0^{-1 }  \star \Gk .
\end{equation}
This implies in particular that
$
	\partial_\epsilon \Gk  =\Gk \star \Big(
	\partial_\epsilon \underline{\Sigma} - \id \Big) \star \Gk$ and $ \nabla_\vec{p} \Gk  = \Gk \star \nabla_\vec{p}H \star \Gk $.
We want to point out that this observation can be used to construct a perturbation series expansion. Assuming we know the exact eigenstates of the Hamiltonian $H_0$ and consider a perturbation $H = H_0 + \lambda \delta H$ controlled by the dimensionless parameter $\lambda$, we can construct the Dyson equation
\begin{equation}
	\Gk= \Gk_{\lambda\to 0}  + \Gk_{\lambda\to0} \star \delta H \star \Gk.
\end{equation}
The perturbation by the electric field is treated fundamentaly differently in our approach as it changes the symplectic structure of the underlying phase-space. In particular, the presence of a non-zero field tensor $F^{\mu\nu}$ implies that
\begin{equation}
 \lim_{F^{\mu\nu} \to 0}	\frac{\partial}{\partial F^{\mu\nu}} \star =  -e\frac{i \hbar}{2}\left(
	  \lpartial_{p^\mu}\rpartial_{p^\nu}
	\right) \circ \star ,
\end{equation}
and is in general a $\mathrm{T}$-symmetry breaking operation.
Applying this derivative to the Dyson equation yields
\begin{align}
 \lim_{F^{\mu\nu} \to 0}	\frac{\partial}{\partial F^{\mu\nu}} \Gk = 
	-e\frac{i \hbar}{2} \Gk \star 
	\partial_{p^\mu} \Gk_0^{-1}\star  \Gk \star   \partial_{p^\nu}  \Gk_0^{-1} \star \Gk .
\end{align}
Since $c F^{i0} = E_i $, a perturbation by a constant, homogeneous electric can be represented by the chain rule
\begin{align}
	\frac{\partial}{\partial E_i} \rightarrow  \frac{1}{c} \left( \frac{\partial}{\partial F^{i0}} 
	 -  \frac{\partial}{\partial F^{0i}} 
	\right) . 
\end{align}
The electric field expansion of the Green's function is then given by $\Gk \to \Gk + \Gk_\vec{E} + \mathcal{O}(E^2)$ , where
\begin{align}
	\Gk_\vec{E} &= -	\frac{i \hbar e}{2} \Big( \Gk \star 
	\nabla_\vec{p} \Gk_0^{-1}\star  \Gk \star   \partial_{\epsilon}  \Gk_0^{-1} \star \Gk 
- \Gk \star \partial_{\epsilon}\Gk_0^{-1}\star  \Gk \star    \nabla_\vec{p}  \Gk_0^{-1} \star \Gk  \Big) \cdot \vec{E}.
\end{align}
The result can be split in two parts: i) a part originating from the Fermi sea and ii) a term originating from the Fermi surface. The first one is given by
\begin{align}
\Gk_{\vec{E}}^\mathrm{sea} &= \frac{i \hbar e}{2} \Big(
\Gk \star \vec{v}  \star  \Gk \star \Gk
-
\Gk\star  \Gk \star \vec{v}  \star \Gk
\Big) \cdot \vec{E},
\end{align}
and it has the property
$
	G_{\vec{E}}^{<, \mathrm{sea}}
	= f(\epsilon) \left( 	\Gk_{\vec{E}}^{\mathrm{A}, \mathrm{sea}} - 	\Gk_{\vec{E}}^{\mathrm{R}, \mathrm{sea}} \right) 
$.
The second contribution to the Green's function is defined by
\begin{align}
G_{\vec{E}}^\mathrm{surf} = -\frac{i \hbar e}{2} \Big(
\Gk \star\vec{v} \star \Gk \star \partial_\epsilon \underline{\Sigma}  \star  \Gk 
-
\Gk \star \partial_\epsilon \underline{\Sigma} \star \Gk \star\vec{v}  \star  \Gk 
\Big) \cdot \vec{E}
\end{align}
Assuming that $\partial_t H = 0$, the Dyson equation leads to the result
$
    \Sigma^< = f(\epsilon)  \left[
        (\Gret_0)^{-1} - (\Gadv_0)^{-1}
    \right] $.
Therefore, the lesser component of the surface term mixes advanced and retarded components:
\begin{align}
  G_{\vec{E}}^{<,\mathrm{surf}}
= -\frac{i \hbar e}{2} \partial_\epsilon f  \Big(
(\Gret - \Gadv) \star  \vec{v} \star \Gadv
- \Gret \star \vec{v} \star (\Gret - \Gadv)
\Big) \cdot \vec{E}.
\end{align}
The important aspect of these to results is the fact that they are of infinite order in the sense of the gradient expansion and non-perturbative in this regard. Only the electromagnetic field is treated in a perturbative way. Inserting both quantities into the expression for the current and taking the derivative $
    \sigma_{kl} = \left.  \partial \Braket{j}_k / \partial E_l \right|_{\vec{E}\to 0}
$ yields the expression which were provided in the manuscript, i.e.,
\begin{align}
    \sigma^\mathrm{sea}_{kl} &= \hbar e^2\Re ~
    \fraktr_\mathrm{sea} \Braket{  \Gret \star v_k \star \Gret  \star v_l \star \Gret - ( k \! \leftrightarrow \! l) },
    \\
     \sigma^\mathrm{surf}_{kl} &= \hbar e^2  \Re ~
    \fraktr_\mathrm{surf} \Braket{
    v_k \star ( \Gadv - \Gret ) \star v_l \star \Gadv 
    }.
    \notag \\
\end{align}

\subsection{Cyclic Hochschild Cohomology}

Here, we briefly review some definitions from cyclic cohomology. The notation follows closely the one in~\cite{Khalkhali2004}. If $A$ is an algebra and $M$ is an $A$-bimodule, one defines the {\it Hochschild cochain} complex of $A$ with coefficients in $M$, denoted by $(C^\bullet(A,M), \delta)$ by
$C^0(A,M) = M$, $C^n(A,M) = \mathrm{Hom}(A^{\otimes n},M)$ for $n\geq 1
$ and with the differential defined as
\begin{align}
     (\delta m)(a_1) &= m a_1- a_1 m
     \notag \\
     (\delta f)(a_1 , \ldots, a_{n+1})
     &= a_1 f(a_2, \ldots, a_{n+1}) + \sum_{i=1}^n (-1)^{i+1} f(a_1, \ldots, a_i a_{i+1}, \ldots, a_{n+1}) + (-1)^{n+1} f(a_1, \ldots , a_n) a_{n+1},
\end{align}
where $m \in C^0$, $f\in C^n$  and $a_i \in A$. It can be shown that $\delta^2 =0$. The cohomology of the Hochschild cochain complex is called the {\it Hochschild cohomology} of $A$ with coefficients in $M$. For the special case where $M$ is dual to $A$, i.e., $M = A^\ast = \mathrm{Hom}(A,k) $ where $k$ is a field of characteristic zero, one can use the fact that
\begin{equation}
    \mathrm{Hom}(A^{\otimes n}, A^\ast) \cong 
    \mathrm{Hom}(A^{\otimes (n+1)}, k),
\end{equation}
to construct a new differential $b$ which operates directly on $\mathrm{Hom}(A^{\otimes (n+1)}, k)$. From now on we use the abbreviated notation $C^n(A) = C^n(A,A^\ast) = \mathrm{Hom}(A^{\otimes (n+1)}, k)$. For $f \in C^n(A)$ we define $b$ by
\begin{align}
    (bf)(a_0,\ldots,a_{n+1}) =
    \sum_{i=0}^n (-1)^i f(a_0, \ldots, a_i a_{i+1}, \ldots , a_{n+1} ) +(-1)^{n+1} f(a_{n+1} a_0, \ldots, a_n).
\end{align}
Let $\lambda: C^n(A) \to C^n(A)$ be given by
\begin{equation}
    (\lambda  f) (a_0, \ldots,a_{n})
    = (-1)^n f(a_n, a_0, \ldots, a_{n-1})
\end{equation}
One then defines the space of cyclic cochains as $C_\lambda^n(A) \equiv \mathrm{ker}(1-\lambda)$. Since $b C_\lambda^n(A) \subset C_\lambda^{n+1}(A) $ one arrives at a cochain complex $(C^\bullet_\lambda(A),b)$ known as the {\it cyclic complex} and which forms a subcomplex of the Hochschild complex defined above. The cohomology of the cyclic complex is the {\it cyclic cohomology} denoted by $HC(A)$, i.e., 
\begin{equation}
    HC^n( A) \equiv \frac{ \mathrm{ker}~ b^{n\phantom{-1} } }{\mathrm{im} ~ b^{n-1}}.
\end{equation}

From now on, we understand $A$ as the non-commutative algebra of matrix valued smooth Keldysh phase space functions w.r.t the Moyal $\star$-product. We then introduce the cochain
\begin{equation}
    \varphi^{2n}(a_0, \ldots , a_{2n} ) 
    \equiv  \Re~\fraktr \Braket{  \epsilon^{\mu_0 \cdots \mu_d} a_{\mu_0}^\mathrm{R} \star a_{\mu_1}^\mathrm{R} \star \ldots \star a_{\mu_{2n}}^\mathrm{R} }.
\end{equation}
From the cyclic property of the phase space trace follows that $\varphi^{2n}$ is a cyclic cochain, i.e., $\varphi^{2n}\in C^{2n}_\lambda(A)$ . The cyclic property also implies that $\varphi^{2n}$ is invariant under inner automorphisms of $A$ which are of the form $a_\mu \to u \star a_\mu \star u^{-1\star}$ for invertible elements $u$ of $A$. Using the cyclic property we rewrite
\begin{align}
      \varphi^{2}(a_0, a_1 , a_{2} ) 
&=  3 \Re~\fraktr \Braket{  a_{0}^\mathrm{R} 
\star
    a_{1}^\mathrm{R} 
    \star 
    a_{2}^\mathrm{R}-a_{2}^\mathrm{R} 
    \star 
    a_{1}^\mathrm{R} 
    \star 
    a_{0}^\mathrm{R}  }
\notag \\
&=  3\Re~\fraktr \Braket{  a_{0}^\mathrm{R} 
\star
    a_{1}^\mathrm{R} 
    \star 
    a_{2}^\mathrm{R}
    - (\mathrm{R} \to \mathrm{A} ) }
\notag \\
&\equiv \psi^{2}_{\mathrm{R}} (a_0, a_1, a_{2} ) - (\mathrm{R} \to \mathrm{A} ).
\label{eq:cocycle_forms}
\end{align}
From the high symmetry of $\psi^{2}$ follows $b\psi^{2}=0$ which extends by linearity to $b\varphi^{2}=0$. Therefore, the cyclic cochain $\varphi^2$ represents a cocycle, i.e, $\varphi^2 \in HC^2(A)$. Due to the invariance under inner automorphism, one might wonder if the Kubo conductivity represents now a pairing
\begin{equation}
    \sigma_{xy}: HC^2(A) \otimes K_0(A) \to \mathbb{R},
\end{equation}
where $K_0(A)$ is the $K$-theory of $A$. However, the Hall-conductivity is not formulated in terms of idempotents (which are a useful ingredient in showing that the pairing depends only on the cyclic cohomology class).

\subsection{Onsager relations}
In the manuscript we defined the Maurer-Cartan like objects
$
    \theta_\mu^\mathrm{R} \equiv \left(\Gret \star \frac{\partial}{\partial p^\mu} (\Gret)^{-1\star } \right)
$. If $\hatn$ represents the local vector field of magnetization, we can express the Kubo cocycle as a functional of $\hatn$, i.e., $ \varphi^{2}(\theta_0, \theta_1 , \theta_{2} )  \equiv \varphi^{2}[\hatn] $. If we now employ operation $\mathrm{T}$ of time-reversal by the relation
$
    \Gret [\hatn] = \mathrm{T} \star \Gadv[-\hatn] \star \mathrm{T}^{-1}
$,
it is apparent from Eq.~(\ref{eq:cocycle_forms}) that this implies
$
    \varphi^{2}[\hatn] = -\varphi^{2}[-\hatn]
$ and as a consequence, the Onsager relation $\sigma_{xy}^\mathrm{sea} [\hatn] = \sigma_{yx}^\mathrm{sea} [-\hatn]
$ holds. In order to derive this, one cannot permute the time-reversal operator under the trace. However, one can use the decomposition of $\mathrm{T}=\mathcal{U} K$ into a unitary operator $\mathcal{U}$ and complex conjugation $K$. Using this decomposition one finds
$
    \Re~ \tr~ \mathrm{T} \mathcal{O} \mathrm{T}^{-1} = \Re~ \tr~ \mathcal{O}^\ast = \Re~ \tr~ \mathcal{O}
$.


\subsection{Topological Invariance}

In the manuscript we defined the Maurer-Cartan like objects
$
    \theta_\mu \equiv \left(\Gret \star \frac{\partial}{\partial p^\mu} (\Gret)^{-1\star } \right)
$ and the cyclic cocycles
\begin{equation}
    \varphi^n(\theta_0, \ldots , \theta_{n} ) 
    \equiv  \fraktr_\mathrm{sea} \Braket{  \epsilon^{\mu_0 \cdots \mu_{n}} \theta_{\mu_0} \star \theta_{\mu_1} \star \ldots \star \theta_{\mu_n} }.
\end{equation}
To proof the topological invariance of $\varphi^{n=d}$ in $d$ space-dimensions under a smooth variation $\delta$ of the retarded Green's function we generalize an argument of~\cite{Qi2008}. We assume that the system is an insulator at zero temperature and the variation does not close the gap. The $\theta$-objects transform as
\begin{equation}
     \delta \theta_\mu
     = - \Gret \star \partial_\mu (\Gret)^{-1\star } \star \delta \Gret \star (\Gret)^{-1\star } - ( \partial_\mu \delta \Gret) \star (\Gret)^{-1\star } =
     - \Gret \star (\partial_\mu \Delta ) \star (\Gret)^{-1\star } ,
\end{equation}
where we have defined $\Delta = (\Gret)^{-1\star } \star \delta \Gret$. Analogously we set $\tilde{\theta}_\mu =(\Gret)^{-1\star } \star \theta_\mu  \star \Gret  = \partial_\mu (\Gret)^{-1\star } \star \Gret  $. The variation of the cocycle is then
\begin{align}
\delta  \varphi^d(\theta_0, \ldots, \theta_d)
& =   \delta \fraktr_\mathrm{sea} \Braket{\epsilon^{\mu_0 \cdots \mu_d} \theta_{\mu_0} \star \theta_{\mu_1} \star \ldots \star \theta_{\mu_d} }
    \notag \\
& \propto   \fraktr_\mathrm{sea} \Braket{\epsilon^{\mu_0 \cdots \mu_d} (\delta \theta_{\mu_0}) \star \theta_{\mu_1} \star \ldots \star \theta_{\mu_d} }
    \notag \\
& =   -\fraktr_\mathrm{sea} \Braket{\epsilon^{\mu_0 \cdots \mu_d} (\partial_{\mu_0} \Delta) \star \tilde{\theta}_{\mu_1} \star \ldots \star \tilde{\theta}_{\mu_d} }
    \notag \\
& \overset{\mathrm{p.i.}}{=}   \fraktr_\mathrm{sea} \Braket{ \Delta \star \epsilon^{\mu_0 \cdots \mu_d} \partial_{\mu_0} (\tilde{\theta}_{\mu_1} \star \ldots \star \tilde{\theta}_{\mu_d} )}.
\end{align}
If the dimension is an even integer, i.e., $d \in 2\mathbb{N}$, cyclic and anti-cyclic terms in $\epsilon^{\mu_0 \cdots \mu_d} \partial_{\mu_0} (\tilde{\theta}_{\mu_1} \star \ldots \star \tilde{\theta}_{\mu_d} )$ pair up and cancel each other, and therefore
\begin{equation}
    \delta  \varphi^d(\theta_0, \ldots, \theta_d) = 0,
\end{equation}
whenever $d \in 2\mathbb{N}$.

\subsection{Ring diagram expansion}

\begin{figure}[t]
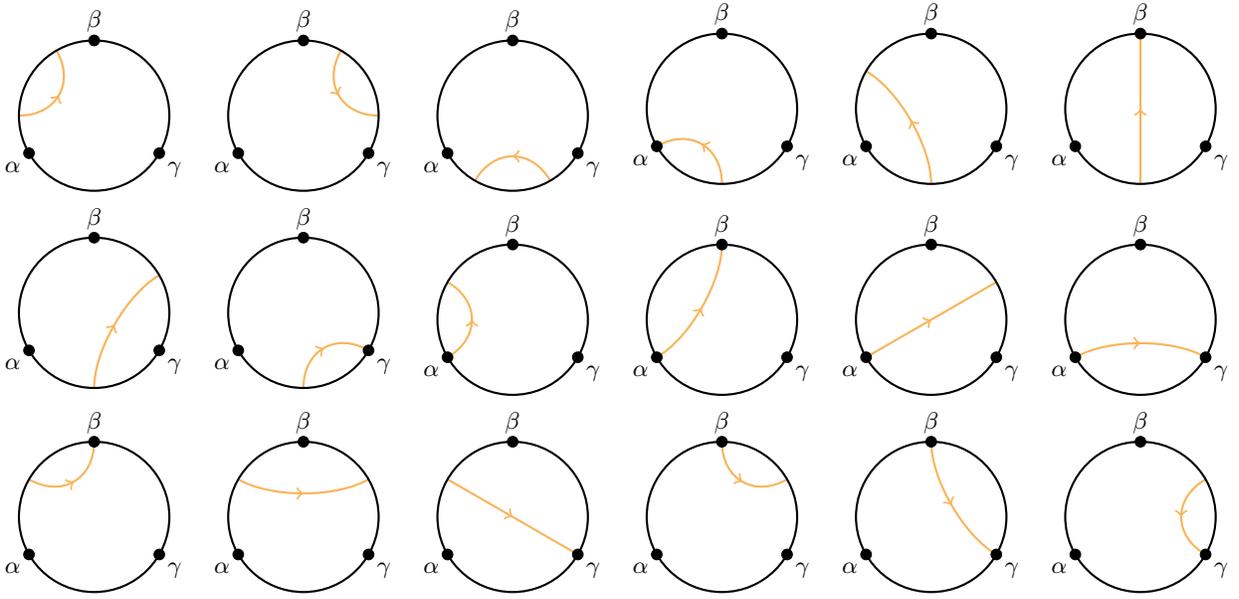


    \centering
    
    \begin{tabular}{c c c c c c}
    	\chediagram{\vertexABl}{\vertexABr} & \chediagram{\vertexBCl}{\vertexBCr} & \chediagram{\vertexCAl}{\vertexCAr} &
    	\chediagram{\vertexCAc}{\vertexA}   &
    	\chediagram{\vertexCAc}{\vertexABc} &
    	\chediagram{\vertexCAc}{\vertexB}   \\
    	\chediagram{\vertexCAc}{\vertexBCc} &
    	\chediagram{\vertexCAc}{\vertexC}   &
    	\chediagram{\vertexA}{\vertexABc}   &
    	\chediagram{\vertexA}{\vertexB}     &
    	\chediagram{\vertexA}{\vertexBCc}   & 
    	\chediagram{\vertexA}{\vertexC}     \\
    	\chediagram{\vertexABc}{\vertexB}   &
    	\chediagram{\vertexABc}{\vertexBCc} & 
    	\chediagram{\vertexABc}{\vertexC}   &
    	\chediagram{\vertexB}{\vertexBCc}   &
    	\chediagram{\vertexB}{\vertexC}     &
    	\chediagram{\vertexBCc}{\vertexC}
    \end{tabular}

    \caption{Diagrammatic expansion of the chiral anomalous Hall effect. The black lines represent retarded Green's functions, the black vertices correspond to four-momentum derivatives of the inverse propagator. Gradient corrections (apricot) have a directionality which is indicated by an arrow. }
    
    \label{fig:che_all_diagrams}
    
\end{figure}

The calculation of the anomalous Hall conductivity requires the evaluation of the $\star$-product chain
\begin{equation}
    \fraktr_\mathrm{occ} \Braket{ \Gret \star  
    \vertex{\alpha}
    \star
     \Gret \star  \vertex{\beta} \star
         \Gret \star  
    \vertex{\gamma}},
\label{eq:kubo_trace}
\end{equation}
where $\vertex{\alpha} = \partial_{p^\alpha} (\Gret)^{-1 \star}$. The zeroth order term can be represented diagrammatically as
\begin{equation}
    \smallkubodiagram .
\end{equation}
The circle captures the cyclic invariance of the trace operation where the solid lines represent retarded Green's functions. In order to translate the diagram into an equation it is to be read in clockwise direction. The next-order correction (the ``one-loop'' level) can be obtained by expanding the Green's functions to first order in the gradients, i.e,
\begin{equation}
    \Gret = \Gret_0 + \frac{i \hbar }{2} \Pi^{ij}
    \Gret_0 \partial_i (\Gret_0)^{-1} \Gret_0 \partial_j (\Gret_0)^{-1}\Gret_0 
\end{equation}
where $\Pi$ represents the symplectic structure of the underlying phase space. Diagrammatically, this correction can be captured as
\begin{equation}
\Gret =   \Gret_0 +  \firstorderdiagram . 
\end{equation}
Inserting this relation to Eq.~\ref{eq:kubo_trace}, expanding the $\star$-products and retaining only first-order contributions in $\hbar$ leads to the set of diagrams presented in Fig.~\ref{fig:che_all_diagrams}. From the associativity of the $\star$-product, one is free to select a certain bracketing, i.e,
\begin{align}
   \Gret \star  
    \vertex{\alpha}
    \star
     \Gret \star  \vertex{\beta} \star
         \Gret \star \vertex{\gamma} 
= \Gret \star  (
    \vertex{\alpha}
    \star (
     \Gret \star  ( \vertex{\beta} \star (
         \Gret \star \vertex{\gamma} )))).
\end{align}

Since we have the identity
$
    \partial_i \Gret_0 = - \Gret_0 \partial_i ( \Gret_0)^{-1} \Gret_0,
$
the diagrams have to multiplied with a factor of $(-1)^{\# v}$, where $\# v$ is the number of velocity vertices $\vertex{\alpha}$, to which the Moyal gradient line is attached. Since the diagrams will be contracted with the Levi-Civita symbol $\epsilon^{\alpha\beta\gamma}$ they are invariant under a cyclic relabelling of $\alpha\beta\gamma$. Making use of this fact, the sum over all diagrammatic contributions simplifies to
\begin{align}
   \Sigma \equiv ~& 3 \smallchediagram{\vertexABl}{\vertexABr}
   + 3 	\smallchediagram{\vertexCAc}{\vertexA}
   + \smallchediagram{\vertexCAc}{\vertexABc}
   \notag \\ &
   +\smallchediagram{\vertexCAc}{\vertexB}
   +\smallchediagram{\vertexA}{\vertexABc} .
\end{align}
Here we have disregarded diagrams of the type
\begin{equation}
    \smallchediagram{\vertexA}{\vertexB} ,
\end{equation}
which are zero for the class of Hamiltonians we consider, i.e., those where position and momentum variables do not couple directly. It can be seen that the sum
\begin{align}
\smallchediagram{\vertexCAc}{\vertexA}
+
   \smallchediagram{\vertexCAc}{\vertexB}
+
\smallchediagram{\vertexCAc}{\vertexC} 
\end{align}
amounts to a total derivative under the trace. By partial integration, we can therefore replace
\begin{align}
    \smallchediagram{\vertexCAc}{\vertexB}
    \rightarrow 
    -
    \smallchediagram{\vertexCAc}{\vertexA}
    -
    \smallchediagram{\vertexCAc}{\vertexC} 
\end{align}
Inserting this relation into the diagrammatic sum yields
\begin{align}
   \Sigma \equiv ~& 3 \smallchediagram{\vertexABl}{\vertexABr}
   + 2 	\smallchediagram{\vertexCAc}{\vertexA}
   + \smallchediagram{\vertexCAc}{\vertexABc}
   + 2\smallchediagram{\vertexA}{\vertexABc} .
\end{align}

The replacement
$
    \vertex{\alpha} \to -\partial_{p_\alpha} \amputatedfirstorderdiagram 
$
would induce the diagrams
\begin{equation}
    \smallchediagram{\vertexCAc}{\vertexA}
    +\smallchediagram{\vertexA}{\vertexABc}
    +\smallchediagram{\vertexCAc}{\vertexABc},
\end{equation}
and is used in the manuscript to compactify the notation.

\clearpage

\section{Spin Dynamics Simulation}

\begin{figure}[h]
    \centering
    \includegraphics[width=0.7\linewidth]{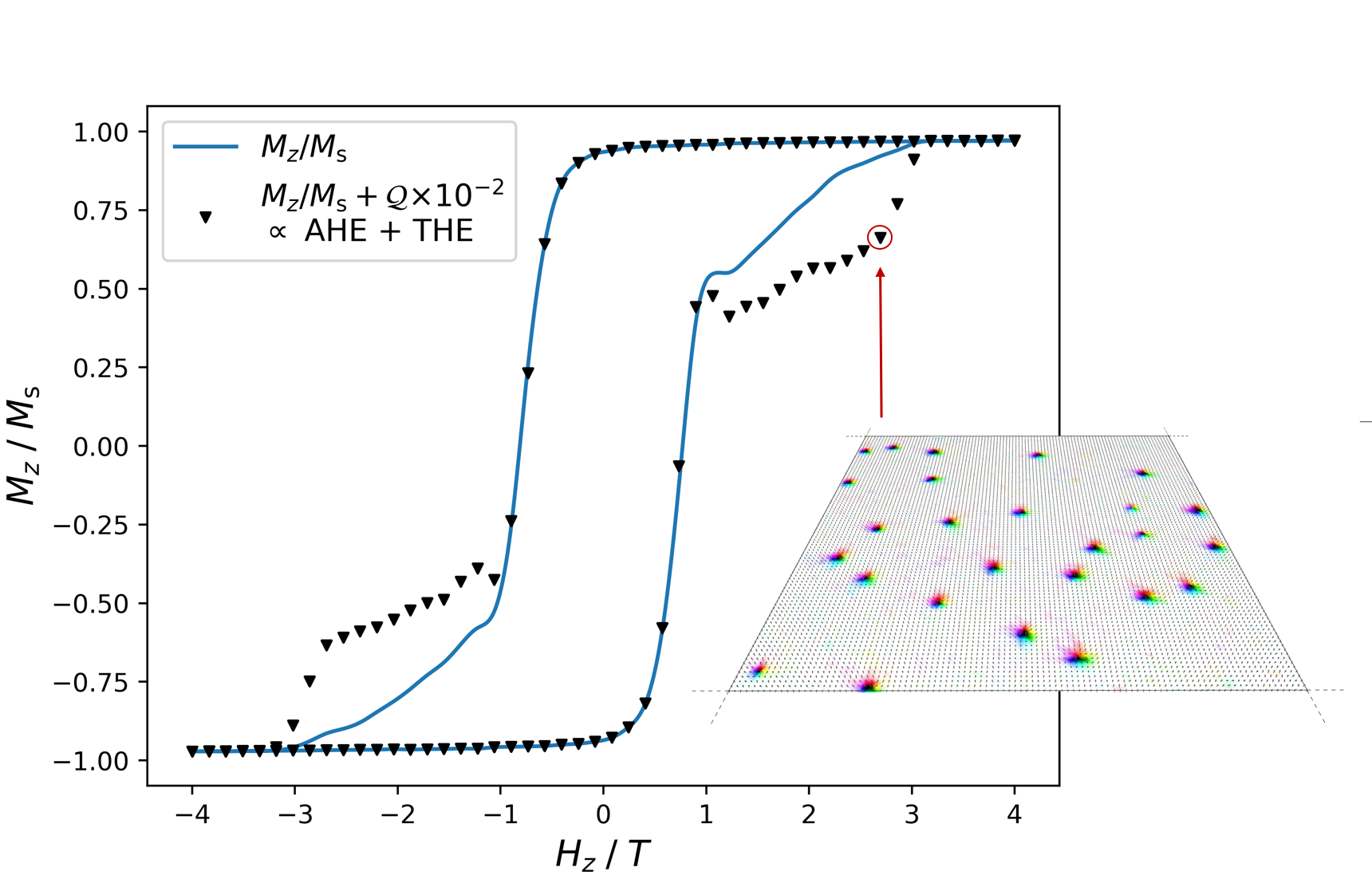}
    \caption{Hysteresis loop for a chiral Heisenberg magnet. The $2D$-system containing $100\times100$ spins is simulated at $T=\SI{1}{\kelvin}$ using the {\it Spirit} code. Upon switching the $H$-field, the systems enters a phase with isolated skyrmions shown in the perspective inset. Measuring the anomalous Hall conductivity along the hysteresis loop could result in a signal proportional to the black data points, where $\mathcal{Q}$ represents the topological charge of the spin structure.}
    \label{fig:full_hysteresis}
\end{figure}

\begin{figure}[ht]
    \centering
    \includegraphics[width=0.7\linewidth]{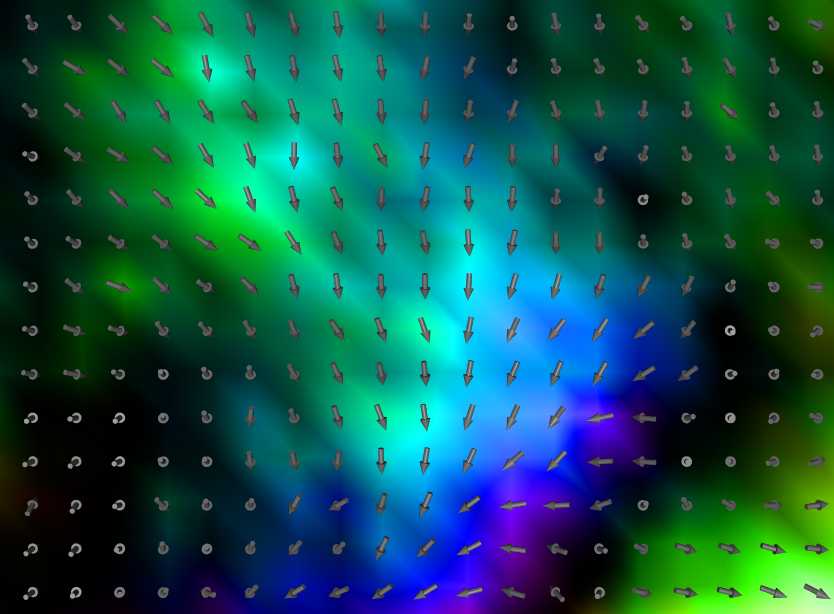}
    \caption{Excerpt from the spin-state at about $\SI{0.57}{\tesla}$ (in the switching branch of the hysteresis loop). This is the local conical phase which was alluded to in the manuscript. The chiral Hall effect shows a strong resonance to these states. The color code represents the in-plane angle of the local spin.}
    \label{fig:conical_phase}
\end{figure}

The spin-dynamics simulation in {\it Spirit}~\cite{Mueller2019} implements a square lattice of $100\times100$ spins (spin-moment $2\mu_\mathrm{B}$) and solves the Landau–Lifshitz–Gilbert equation using a Depondt solver at $T=\SI{1}{\kelvin}$. The atomistic Hamiltonian consists of the isotropic nearest-neighbor Heisenberg exchange $J = 1 \SI{1}{\milli\electronvolt}$ and the Dzyaloshinskii–Moriya interaction of $D = 0.5 \SI{1}{\milli\electronvolt}$. This system stabilizes skyrmions in between $\SI{2}{\tesla}$ and $\SI{3}{\tesla}$. To perform a hysteresis loop experiment starting from a polarized state, the magnetic field is reduced in steps of about $\SI{0.16}{\tesla}$. After each step, the system evolves for $\SI{8}{\pico\second}$ before the next increment in the magnetic field strength occurs. At $\SI{4}{\tesla}$, the system is fully polarized and the magnetic field is swept in the opposite direction until the loop is complete. The result is shown in Fig.~\ref{fig:full_hysteresis} with an excerpt from the spin state at $\SI{0.57}{\tesla}$ in Fig.~\ref{fig:conical_phase}.

\begin{thebibliography}{43}%
\makeatletter
\providecommand \@ifxundefined [1]{%
 \@ifx{#1\undefined}
}%
\providecommand \@ifnum [1]{%
 \ifnum #1\expandafter \@firstoftwo
 \else \expandafter \@secondoftwo
 \fi
}%
\providecommand \@ifx [1]{%
 \ifx #1\expandafter \@firstoftwo
 \else \expandafter \@secondoftwo
 \fi
}%
\providecommand \natexlab [1]{#1}%
\providecommand \enquote  [1]{``#1''}%
\providecommand \bibnamefont  [1]{#1}%
\providecommand \bibfnamefont [1]{#1}%
\providecommand \citenamefont [1]{#1}%
\providecommand \href@noop [0]{\@secondoftwo}%
\providecommand \href [0]{\begingroup \@sanitize@url \@href}%
\providecommand \@href[1]{\@@startlink{#1}\@@href}%
\providecommand \@@href[1]{\endgroup#1\@@endlink}%
\providecommand \@sanitize@url [0]{\catcode `\\12\catcode `\$12\catcode
  `\&12\catcode `\#12\catcode `\^12\catcode `\_12\catcode `\%12\relax}%
\providecommand \@@startlink[1]{}%
\providecommand \@@endlink[0]{}%
\providecommand \url  [0]{\begingroup\@sanitize@url \@url }%
\providecommand \@url [1]{\endgroup\@href {#1}{\urlprefix }}%
\providecommand \urlprefix  [0]{URL }%
\providecommand \Eprint [0]{\href }%
\providecommand \doibase [0]{https://doi.org/}%
\providecommand \selectlanguage [0]{\@gobble}%
\providecommand \bibinfo  [0]{\@secondoftwo}%
\providecommand \bibfield  [0]{\@secondoftwo}%
\providecommand \translation [1]{[#1]}%
\providecommand \BibitemOpen [0]{}%
\providecommand \bibitemStop [0]{}%
\providecommand \bibitemNoStop [0]{.\EOS\space}%
\providecommand \EOS [0]{\spacefactor3000\relax}%
\providecommand \BibitemShut  [1]{\csname bibitem#1\endcsname}%
\let\auto@bib@innerbib\@empty
\bibitem [{\citenamefont {Fert}\ \emph {et~al.}(2017)\citenamefont {Fert},
  \citenamefont {Reyren},\ and\ \citenamefont {Cros}}]{Fert2017}%
  \BibitemOpen
  \bibfield  {author} {\bibinfo {author} {\bibfnamefont {A.}~\bibnamefont
  {Fert}}, \bibinfo {author} {\bibfnamefont {N.}~\bibnamefont {Reyren}},\ and\
  \bibinfo {author} {\bibfnamefont {V.}~\bibnamefont {Cros}},\ }\href@noop {}
  {\bibfield  {journal} {\bibinfo  {journal} {Nat. Rev. Mater.}\ }\textbf
  {\bibinfo {volume} {2}},\ \bibinfo {pages} {17031} (\bibinfo {year}
  {2017})}\BibitemShut {NoStop}%
\bibitem [{\citenamefont {Kang}\ \emph {et~al.}(2016)\citenamefont {Kang},
  \citenamefont {Huang}, \citenamefont {Zhang}, \citenamefont {Zhou},\ and\
  \citenamefont {Zhao}}]{Kang2016}%
  \BibitemOpen
  \bibfield  {author} {\bibinfo {author} {\bibfnamefont {W.}~\bibnamefont
  {Kang}}, \bibinfo {author} {\bibfnamefont {Y.}~\bibnamefont {Huang}},
  \bibinfo {author} {\bibfnamefont {X.}~\bibnamefont {Zhang}}, \bibinfo
  {author} {\bibfnamefont {Y.}~\bibnamefont {Zhou}},\ and\ \bibinfo {author}
  {\bibfnamefont {W.}~\bibnamefont {Zhao}},\ }\href@noop {} {\bibfield
  {journal} {\bibinfo  {journal} {Proc. IEEE}\ }\textbf {\bibinfo {volume}
  {104}},\ \bibinfo {pages} {2040} (\bibinfo {year} {2016})}\BibitemShut
  {NoStop}%
\bibitem [{\citenamefont {Woo}\ \emph {et~al.}(2016)\citenamefont {Woo},
  \citenamefont {Litzius}, \citenamefont {Krüger}, \citenamefont {Im},
  \citenamefont {Caretta}, \citenamefont {Richter}, \citenamefont {Mann},
  \citenamefont {Krone}, \citenamefont {Reeve}, \citenamefont {Weigand},
  \citenamefont {Agrawal}, \citenamefont {Lemesh}, \citenamefont {Mawass},
  \citenamefont {Fischer}, \citenamefont {Kläui},\ and\ \citenamefont
  {Beach}}]{Woo2016}%
  \BibitemOpen
  \bibfield  {author} {\bibinfo {author} {\bibfnamefont {S.}~\bibnamefont
  {Woo}}, \bibinfo {author} {\bibfnamefont {K.}~\bibnamefont {Litzius}},
  \bibinfo {author} {\bibfnamefont {B.}~\bibnamefont {Krüger}}, \bibinfo
  {author} {\bibfnamefont {M.-Y.}\ \bibnamefont {Im}}, \bibinfo {author}
  {\bibfnamefont {L.}~\bibnamefont {Caretta}}, \bibinfo {author} {\bibfnamefont
  {K.}~\bibnamefont {Richter}}, \bibinfo {author} {\bibfnamefont
  {M.}~\bibnamefont {Mann}}, \bibinfo {author} {\bibfnamefont {A.}~\bibnamefont
  {Krone}}, \bibinfo {author} {\bibfnamefont {R.~M.}\ \bibnamefont {Reeve}},
  \bibinfo {author} {\bibfnamefont {M.}~\bibnamefont {Weigand}}, \bibinfo
  {author} {\bibfnamefont {P.}~\bibnamefont {Agrawal}}, \bibinfo {author}
  {\bibfnamefont {I.}~\bibnamefont {Lemesh}}, \bibinfo {author} {\bibfnamefont
  {M.-A.}\ \bibnamefont {Mawass}}, \bibinfo {author} {\bibfnamefont
  {P.}~\bibnamefont {Fischer}}, \bibinfo {author} {\bibfnamefont
  {M.}~\bibnamefont {Kläui}},\ and\ \bibinfo {author} {\bibfnamefont
  {G.~S.~D.}\ \bibnamefont {Beach}},\ }\href {https://doi.org/10.1038/nmat4593}
  {\bibfield  {journal} {\bibinfo  {journal} {Nat. Mater.}\ }\textbf {\bibinfo
  {volume} {15}},\ \bibinfo {pages} {501} (\bibinfo {year} {2016})}\BibitemShut
  {NoStop}%
\bibitem [{\citenamefont {Legrand}\ \emph {et~al.}(2019)\citenamefont
  {Legrand}, \citenamefont {Maccariello}, \citenamefont {Ajejas}, \citenamefont
  {Collin}, \citenamefont {Vecchiola}, \citenamefont {Bouzehouane},
  \citenamefont {Reyren}, \citenamefont {Cros},\ and\ \citenamefont
  {Fert}}]{Legrand2019}%
  \BibitemOpen
  \bibfield  {author} {\bibinfo {author} {\bibfnamefont {W.}~\bibnamefont
  {Legrand}}, \bibinfo {author} {\bibfnamefont {D.}~\bibnamefont
  {Maccariello}}, \bibinfo {author} {\bibfnamefont {F.}~\bibnamefont {Ajejas}},
  \bibinfo {author} {\bibfnamefont {S.}~\bibnamefont {Collin}}, \bibinfo
  {author} {\bibfnamefont {A.}~\bibnamefont {Vecchiola}}, \bibinfo {author}
  {\bibfnamefont {K.}~\bibnamefont {Bouzehouane}}, \bibinfo {author}
  {\bibfnamefont {N.}~\bibnamefont {Reyren}}, \bibinfo {author} {\bibfnamefont
  {V.}~\bibnamefont {Cros}},\ and\ \bibinfo {author} {\bibfnamefont
  {A.}~\bibnamefont {Fert}},\ }\href
  {https://doi.org/10.1038/s41563-019-0468-3} {\bibfield  {journal} {\bibinfo
  {journal} {Nat. Mater.}\ } (\bibinfo {year} {2019})}\BibitemShut {NoStop}%
\bibitem [{\citenamefont {Litzius}\ \emph {et~al.}(2017)\citenamefont
  {Litzius}, \citenamefont {Lemesh}, \citenamefont {Kr{\"u}ger}, \citenamefont
  {Bassirian}, \citenamefont {Caretta}, \citenamefont {Richter}, \citenamefont
  {B{\"u}ttner}, \citenamefont {Sato}, \citenamefont {Tretiakov}, \citenamefont
  {F{\"o}rster} \emph {et~al.}}]{Litzius2017}%
  \BibitemOpen
  \bibfield  {author} {\bibinfo {author} {\bibfnamefont {K.}~\bibnamefont
  {Litzius}}, \bibinfo {author} {\bibfnamefont {I.}~\bibnamefont {Lemesh}},
  \bibinfo {author} {\bibfnamefont {B.}~\bibnamefont {Kr{\"u}ger}}, \bibinfo
  {author} {\bibfnamefont {P.}~\bibnamefont {Bassirian}}, \bibinfo {author}
  {\bibfnamefont {L.}~\bibnamefont {Caretta}}, \bibinfo {author} {\bibfnamefont
  {K.}~\bibnamefont {Richter}}, \bibinfo {author} {\bibfnamefont
  {F.}~\bibnamefont {B{\"u}ttner}}, \bibinfo {author} {\bibfnamefont
  {K.}~\bibnamefont {Sato}}, \bibinfo {author} {\bibfnamefont {O.~A.}\
  \bibnamefont {Tretiakov}}, \bibinfo {author} {\bibfnamefont {J.}~\bibnamefont
  {F{\"o}rster}}, \emph {et~al.},\ }\href@noop {} {\bibfield  {journal}
  {\bibinfo  {journal} {Nat. Phys.}\ }\textbf {\bibinfo {volume} {13}},\
  \bibinfo {pages} {170} (\bibinfo {year} {2017})}\BibitemShut {NoStop}%
\bibitem [{\citenamefont {Jiang}\ \emph {et~al.}(2017)\citenamefont {Jiang},
  \citenamefont {Zhang}, \citenamefont {Yu}, \citenamefont {Zhang},
  \citenamefont {Wang}, \citenamefont {Jungfleisch}, \citenamefont {Pearson},
  \citenamefont {Cheng}, \citenamefont {Heinonen}, \citenamefont {Wang} \emph
  {et~al.}}]{Jiang2017}%
  \BibitemOpen
  \bibfield  {author} {\bibinfo {author} {\bibfnamefont {W.}~\bibnamefont
  {Jiang}}, \bibinfo {author} {\bibfnamefont {X.}~\bibnamefont {Zhang}},
  \bibinfo {author} {\bibfnamefont {G.}~\bibnamefont {Yu}}, \bibinfo {author}
  {\bibfnamefont {W.}~\bibnamefont {Zhang}}, \bibinfo {author} {\bibfnamefont
  {X.}~\bibnamefont {Wang}}, \bibinfo {author} {\bibfnamefont {M.~B.}\
  \bibnamefont {Jungfleisch}}, \bibinfo {author} {\bibfnamefont {J.~E.}\
  \bibnamefont {Pearson}}, \bibinfo {author} {\bibfnamefont {X.}~\bibnamefont
  {Cheng}}, \bibinfo {author} {\bibfnamefont {O.}~\bibnamefont {Heinonen}},
  \bibinfo {author} {\bibfnamefont {K.~L.}\ \bibnamefont {Wang}}, \emph
  {et~al.},\ }\href@noop {} {\bibfield  {journal} {\bibinfo  {journal} {Nat.
  Phys.}\ }\textbf {\bibinfo {volume} {13}},\ \bibinfo {pages} {162} (\bibinfo
  {year} {2017})}\BibitemShut {NoStop}%
\bibitem [{\citenamefont {Hamamoto}\ \emph {et~al.}(2016)\citenamefont
  {Hamamoto}, \citenamefont {Ezawa},\ and\ \citenamefont
  {Nagaosa}}]{Hamamoto2016}%
  \BibitemOpen
  \bibfield  {author} {\bibinfo {author} {\bibfnamefont {K.}~\bibnamefont
  {Hamamoto}}, \bibinfo {author} {\bibfnamefont {M.}~\bibnamefont {Ezawa}},\
  and\ \bibinfo {author} {\bibfnamefont {N.}~\bibnamefont {Nagaosa}},\
  }\href@noop {} {\bibfield  {journal} {\bibinfo  {journal} {Appl. Phys.
  Lett.}\ }\textbf {\bibinfo {volume} {108}},\ \bibinfo {pages} {112401}
  (\bibinfo {year} {2016})}\BibitemShut {NoStop}%
\bibitem [{\citenamefont {Maccariello}\ \emph {et~al.}(2018)\citenamefont
  {Maccariello}, \citenamefont {Legrand}, \citenamefont {Reyren}, \citenamefont
  {Garcia}, \citenamefont {Bouzehouane}, \citenamefont {Collin}, \citenamefont
  {Cros},\ and\ \citenamefont {Fert}}]{Maccariello2018}%
  \BibitemOpen
  \bibfield  {author} {\bibinfo {author} {\bibfnamefont {D.}~\bibnamefont
  {Maccariello}}, \bibinfo {author} {\bibfnamefont {W.}~\bibnamefont
  {Legrand}}, \bibinfo {author} {\bibfnamefont {N.}~\bibnamefont {Reyren}},
  \bibinfo {author} {\bibfnamefont {K.}~\bibnamefont {Garcia}}, \bibinfo
  {author} {\bibfnamefont {K.}~\bibnamefont {Bouzehouane}}, \bibinfo {author}
  {\bibfnamefont {S.}~\bibnamefont {Collin}}, \bibinfo {author} {\bibfnamefont
  {V.}~\bibnamefont {Cros}},\ and\ \bibinfo {author} {\bibfnamefont
  {A.}~\bibnamefont {Fert}},\ }\href@noop {} {\bibfield  {journal} {\bibinfo
  {journal} {Nat. Nanotechnol.}\ }\textbf {\bibinfo {volume} {13}},\ \bibinfo
  {pages} {233} (\bibinfo {year} {2018})}\BibitemShut {NoStop}%
\bibitem [{\citenamefont {Matsuno}\ \emph {et~al.}(2016)\citenamefont
  {Matsuno}, \citenamefont {Ogawa}, \citenamefont {Yasuda}, \citenamefont
  {Kagawa}, \citenamefont {Koshibae}, \citenamefont {Nagaosa}, \citenamefont
  {Tokura},\ and\ \citenamefont {Kawasaki}}]{Matsuno2016}%
  \BibitemOpen
  \bibfield  {author} {\bibinfo {author} {\bibfnamefont {J.}~\bibnamefont
  {Matsuno}}, \bibinfo {author} {\bibfnamefont {N.}~\bibnamefont {Ogawa}},
  \bibinfo {author} {\bibfnamefont {K.}~\bibnamefont {Yasuda}}, \bibinfo
  {author} {\bibfnamefont {F.}~\bibnamefont {Kagawa}}, \bibinfo {author}
  {\bibfnamefont {W.}~\bibnamefont {Koshibae}}, \bibinfo {author}
  {\bibfnamefont {N.}~\bibnamefont {Nagaosa}}, \bibinfo {author} {\bibfnamefont
  {Y.}~\bibnamefont {Tokura}},\ and\ \bibinfo {author} {\bibfnamefont
  {M.}~\bibnamefont {Kawasaki}},\ }\href@noop {} {\bibfield  {journal}
  {\bibinfo  {journal} {Sci. Adv.}\ }\textbf {\bibinfo {volume} {2}},\ \bibinfo
  {pages} {e1600304} (\bibinfo {year} {2016})}\BibitemShut {NoStop}%
\bibitem [{\citenamefont {Ohuchi}\ \emph {et~al.}(2018)\citenamefont {Ohuchi},
  \citenamefont {Matsuno}, \citenamefont {Ogawa}, \citenamefont {Kozuka},
  \citenamefont {Uchida}, \citenamefont {Tokura},\ and\ \citenamefont
  {Kawasaki}}]{Ohuchi2018}%
  \BibitemOpen
  \bibfield  {author} {\bibinfo {author} {\bibfnamefont {Y.}~\bibnamefont
  {Ohuchi}}, \bibinfo {author} {\bibfnamefont {J.}~\bibnamefont {Matsuno}},
  \bibinfo {author} {\bibfnamefont {N.}~\bibnamefont {Ogawa}}, \bibinfo
  {author} {\bibfnamefont {Y.}~\bibnamefont {Kozuka}}, \bibinfo {author}
  {\bibfnamefont {M.}~\bibnamefont {Uchida}}, \bibinfo {author} {\bibfnamefont
  {Y.}~\bibnamefont {Tokura}},\ and\ \bibinfo {author} {\bibfnamefont
  {M.}~\bibnamefont {Kawasaki}},\ }\href@noop {} {\bibfield  {journal}
  {\bibinfo  {journal} {Nat. Commun.}\ }\textbf {\bibinfo {volume} {9}},\
  \bibinfo {pages} {213} (\bibinfo {year} {2018})}\BibitemShut {NoStop}%
\bibitem [{\citenamefont {Meng}\ \emph {et~al.}(2019)\citenamefont {Meng},
  \citenamefont {Ahmed}, \citenamefont {Baćani}, \citenamefont {Mandru},
  \citenamefont {Zhao}, \citenamefont {Bagués}, \citenamefont {Esser},
  \citenamefont {Flores}, \citenamefont {McComb}, \citenamefont {Hug} \emph
  {et~al.}}]{Meng2019}%
  \BibitemOpen
  \bibfield  {author} {\bibinfo {author} {\bibfnamefont {K.-Y.}\ \bibnamefont
  {Meng}}, \bibinfo {author} {\bibfnamefont {A.~S.}\ \bibnamefont {Ahmed}},
  \bibinfo {author} {\bibfnamefont {M.}~\bibnamefont {Baćani}}, \bibinfo
  {author} {\bibfnamefont {A.-O.}\ \bibnamefont {Mandru}}, \bibinfo {author}
  {\bibfnamefont {X.}~\bibnamefont {Zhao}}, \bibinfo {author} {\bibfnamefont
  {N.}~\bibnamefont {Bagués}}, \bibinfo {author} {\bibfnamefont {B.~D.}\
  \bibnamefont {Esser}}, \bibinfo {author} {\bibfnamefont {J.}~\bibnamefont
  {Flores}}, \bibinfo {author} {\bibfnamefont {D.~W.}\ \bibnamefont {McComb}},
  \bibinfo {author} {\bibfnamefont {H.~J.}\ \bibnamefont {Hug}}, \emph
  {et~al.},\ }\href@noop {} {\bibfield  {journal} {\bibinfo  {journal} {Nano
  Lett.}\ }\textbf {\bibinfo {volume} {19}},\ \bibinfo {pages} {3169} (\bibinfo
  {year} {2019})}\BibitemShut {NoStop}%
\bibitem [{\citenamefont {Bruno}\ \emph {et~al.}(2004)\citenamefont {Bruno},
  \citenamefont {Dugaev},\ and\ \citenamefont {Taillefumier}}]{Bruno2004}%
  \BibitemOpen
  \bibfield  {author} {\bibinfo {author} {\bibfnamefont {P.}~\bibnamefont
  {Bruno}}, \bibinfo {author} {\bibfnamefont {V.}~\bibnamefont {Dugaev}},\ and\
  \bibinfo {author} {\bibfnamefont {M.}~\bibnamefont {Taillefumier}},\
  }\href@noop {} {\bibfield  {journal} {\bibinfo  {journal} {Phys. Rev. Lett.}\
  }\textbf {\bibinfo {volume} {93}},\ \bibinfo {pages} {096806} (\bibinfo
  {year} {2004})}\BibitemShut {NoStop}%
\bibitem [{\citenamefont {Bliokh}\ and\ \citenamefont
  {Bliokh}(2005)}]{Bliokh2005}%
  \BibitemOpen
  \bibfield  {author} {\bibinfo {author} {\bibfnamefont {K.~Y.}\ \bibnamefont
  {Bliokh}}\ and\ \bibinfo {author} {\bibfnamefont {Y.~P.}\ \bibnamefont
  {Bliokh}},\ }\href@noop {} {\bibfield  {journal} {\bibinfo  {journal} {Ann.
  Phys.}\ }\textbf {\bibinfo {volume} {319}},\ \bibinfo {pages} {13} (\bibinfo
  {year} {2005})}\BibitemShut {NoStop}%
\bibitem [{\citenamefont {Franz}\ \emph {et~al.}(2014)\citenamefont {Franz},
  \citenamefont {Freimuth}, \citenamefont {Bauer}, \citenamefont {Ritz},
  \citenamefont {Schnarr}, \citenamefont {Duvinage}, \citenamefont {Adams},
  \citenamefont {Bl{\"u}gel}, \citenamefont {Rosch}, \citenamefont {Mokrousov}
  \emph {et~al.}}]{Franz2014}%
  \BibitemOpen
  \bibfield  {author} {\bibinfo {author} {\bibfnamefont {C.}~\bibnamefont
  {Franz}}, \bibinfo {author} {\bibfnamefont {F.}~\bibnamefont {Freimuth}},
  \bibinfo {author} {\bibfnamefont {A.}~\bibnamefont {Bauer}}, \bibinfo
  {author} {\bibfnamefont {R.}~\bibnamefont {Ritz}}, \bibinfo {author}
  {\bibfnamefont {C.}~\bibnamefont {Schnarr}}, \bibinfo {author} {\bibfnamefont
  {C.}~\bibnamefont {Duvinage}}, \bibinfo {author} {\bibfnamefont
  {T.}~\bibnamefont {Adams}}, \bibinfo {author} {\bibfnamefont
  {S.}~\bibnamefont {Bl{\"u}gel}}, \bibinfo {author} {\bibfnamefont
  {A.}~\bibnamefont {Rosch}}, \bibinfo {author} {\bibfnamefont
  {Y.}~\bibnamefont {Mokrousov}}, \emph {et~al.},\ }\href@noop {} {\bibfield
  {journal} {\bibinfo  {journal} {Phys. Rev. Lett.}\ }\textbf {\bibinfo
  {volume} {112}},\ \bibinfo {pages} {186601} (\bibinfo {year}
  {2014})}\BibitemShut {NoStop}%
\bibitem [{\citenamefont {Neubauer}\ \emph {et~al.}(2009)\citenamefont
  {Neubauer}, \citenamefont {Pfleiderer}, \citenamefont {Binz}, \citenamefont
  {Rosch}, \citenamefont {Ritz}, \citenamefont {Niklowitz},\ and\ \citenamefont
  {B{\"o}ni}}]{Neubauer2009}%
  \BibitemOpen
  \bibfield  {author} {\bibinfo {author} {\bibfnamefont {A.}~\bibnamefont
  {Neubauer}}, \bibinfo {author} {\bibfnamefont {C.}~\bibnamefont
  {Pfleiderer}}, \bibinfo {author} {\bibfnamefont {B.}~\bibnamefont {Binz}},
  \bibinfo {author} {\bibfnamefont {A.}~\bibnamefont {Rosch}}, \bibinfo
  {author} {\bibfnamefont {R.}~\bibnamefont {Ritz}}, \bibinfo {author}
  {\bibfnamefont {P.}~\bibnamefont {Niklowitz}},\ and\ \bibinfo {author}
  {\bibfnamefont {P.}~\bibnamefont {B{\"o}ni}},\ }\href@noop {} {\bibfield
  {journal} {\bibinfo  {journal} {Phys. Rev. Lett.}\ }\textbf {\bibinfo
  {volume} {102}},\ \bibinfo {pages} {186602} (\bibinfo {year}
  {2009})}\BibitemShut {NoStop}%
\bibitem [{\citenamefont {Nakazawa}\ \emph {et~al.}(2018)\citenamefont
  {Nakazawa}, \citenamefont {Bibes},\ and\ \citenamefont
  {Kohno}}]{Nakazawa2018}%
  \BibitemOpen
  \bibfield  {author} {\bibinfo {author} {\bibfnamefont {K.}~\bibnamefont
  {Nakazawa}}, \bibinfo {author} {\bibfnamefont {M.}~\bibnamefont {Bibes}},\
  and\ \bibinfo {author} {\bibfnamefont {H.}~\bibnamefont {Kohno}},\
  }\href@noop {} {\bibfield  {journal} {\bibinfo  {journal} {J. Phys. Soc.
  Jpn.}\ }\textbf {\bibinfo {volume} {87}},\ \bibinfo {pages} {033705}
  (\bibinfo {year} {2018})}\BibitemShut {NoStop}%
\bibitem [{\citenamefont {Vistoli}\ \emph {et~al.}(2019)\citenamefont
  {Vistoli}, \citenamefont {Wang}, \citenamefont {Sander}, \citenamefont {Zhu},
  \citenamefont {Casals}, \citenamefont {Cichelero}, \citenamefont
  {Barth{\'e}l{\'e}my}, \citenamefont {Fusil}, \citenamefont {Herranz},
  \citenamefont {Valencia} \emph {et~al.}}]{Vistoli2019}%
  \BibitemOpen
  \bibfield  {author} {\bibinfo {author} {\bibfnamefont {L.}~\bibnamefont
  {Vistoli}}, \bibinfo {author} {\bibfnamefont {W.}~\bibnamefont {Wang}},
  \bibinfo {author} {\bibfnamefont {A.}~\bibnamefont {Sander}}, \bibinfo
  {author} {\bibfnamefont {Q.}~\bibnamefont {Zhu}}, \bibinfo {author}
  {\bibfnamefont {B.}~\bibnamefont {Casals}}, \bibinfo {author} {\bibfnamefont
  {R.}~\bibnamefont {Cichelero}}, \bibinfo {author} {\bibfnamefont
  {A.}~\bibnamefont {Barth{\'e}l{\'e}my}}, \bibinfo {author} {\bibfnamefont
  {S.}~\bibnamefont {Fusil}}, \bibinfo {author} {\bibfnamefont
  {G.}~\bibnamefont {Herranz}}, \bibinfo {author} {\bibfnamefont
  {S.}~\bibnamefont {Valencia}}, \emph {et~al.},\ }\href@noop {} {\bibfield
  {journal} {\bibinfo  {journal} {Nat. Phys.}\ }\textbf {\bibinfo {volume}
  {15}},\ \bibinfo {pages} {67} (\bibinfo {year} {2019})}\BibitemShut {NoStop}%
\bibitem [{\citenamefont {Denisov}\ \emph {et~al.}(2018)\citenamefont
  {Denisov}, \citenamefont {Rozhansky}, \citenamefont {Averkiev},\ and\
  \citenamefont {L{\"a}hderanta}}]{Denisov2018}%
  \BibitemOpen
  \bibfield  {author} {\bibinfo {author} {\bibfnamefont {K.}~\bibnamefont
  {Denisov}}, \bibinfo {author} {\bibfnamefont {I.}~\bibnamefont {Rozhansky}},
  \bibinfo {author} {\bibfnamefont {N.}~\bibnamefont {Averkiev}},\ and\
  \bibinfo {author} {\bibfnamefont {E.}~\bibnamefont {L{\"a}hderanta}},\
  }\href@noop {} {\bibfield  {journal} {\bibinfo  {journal} {Phys. Rev. B}\
  }\textbf {\bibinfo {volume} {98}},\ \bibinfo {pages} {195439} (\bibinfo
  {year} {2018})}\BibitemShut {NoStop}%
\bibitem [{\citenamefont {Rozhansky}\ \emph {et~al.}(2019)\citenamefont
  {Rozhansky}, \citenamefont {Denisov}, \citenamefont {Lifshits}, \citenamefont
  {Averkiev},\ and\ \citenamefont {L{\"a}hderanta}}]{Rozhansky2019}%
  \BibitemOpen
  \bibfield  {author} {\bibinfo {author} {\bibfnamefont {I.}~\bibnamefont
  {Rozhansky}}, \bibinfo {author} {\bibfnamefont {K.}~\bibnamefont {Denisov}},
  \bibinfo {author} {\bibfnamefont {M.}~\bibnamefont {Lifshits}}, \bibinfo
  {author} {\bibfnamefont {N.}~\bibnamefont {Averkiev}},\ and\ \bibinfo
  {author} {\bibfnamefont {E.}~\bibnamefont {L{\"a}hderanta}},\ }\href@noop {}
  {\bibfield  {journal} {\bibinfo  {journal} {Phys. Status Solidi B}\ ,\
  \bibinfo {pages} {1900033}} (\bibinfo {year} {2019})}\BibitemShut {NoStop}%
\bibitem [{\citenamefont {Nakabayashi}\ and\ \citenamefont
  {Tatara}(2014)}]{Nakabayashi2014}%
  \BibitemOpen
  \bibfield  {author} {\bibinfo {author} {\bibfnamefont {N.}~\bibnamefont
  {Nakabayashi}}\ and\ \bibinfo {author} {\bibfnamefont {G.}~\bibnamefont
  {Tatara}},\ }\href@noop {} {\bibfield  {journal} {\bibinfo  {journal} {New J.
  Phys.}\ }\textbf {\bibinfo {volume} {16}},\ \bibinfo {pages} {015016}
  (\bibinfo {year} {2014})}\BibitemShut {NoStop}%
\bibitem [{\citenamefont {Lux}\ \emph {et~al.}(2018)\citenamefont {Lux},
  \citenamefont {Freimuth}, \citenamefont {Bl{\"u}gel},\ and\ \citenamefont
  {Mokrousov}}]{Lux2018}%
  \BibitemOpen
  \bibfield  {author} {\bibinfo {author} {\bibfnamefont {F.~R.}\ \bibnamefont
  {Lux}}, \bibinfo {author} {\bibfnamefont {F.}~\bibnamefont {Freimuth}},
  \bibinfo {author} {\bibfnamefont {S.}~\bibnamefont {Bl{\"u}gel}},\ and\
  \bibinfo {author} {\bibfnamefont {Y.}~\bibnamefont {Mokrousov}},\ }\href@noop
  {} {\bibfield  {journal} {\bibinfo  {journal} {Commun. Phys.}\ }\textbf
  {\bibinfo {volume} {1}},\ \bibinfo {pages} {60} (\bibinfo {year}
  {2018})}\BibitemShut {NoStop}%
\bibitem [{\citenamefont {Onoda}\ \emph {et~al.}(2006)\citenamefont {Onoda},
  \citenamefont {Sugimoto},\ and\ \citenamefont {Nagaosa}}]{Onoda2006}%
  \BibitemOpen
  \bibfield  {author} {\bibinfo {author} {\bibfnamefont {S.}~\bibnamefont
  {Onoda}}, \bibinfo {author} {\bibfnamefont {N.}~\bibnamefont {Sugimoto}},\
  and\ \bibinfo {author} {\bibfnamefont {N.}~\bibnamefont {Nagaosa}},\ }\href
  {https://doi.org/10.1143/PTP.116.61} {\bibfield  {journal} {\bibinfo
  {journal} {Prog. Theor. Phys.}\ }\textbf {\bibinfo {volume} {116}},\ \bibinfo
  {pages} {61} (\bibinfo {year} {2006})}\BibitemShut {NoStop}%
\bibitem [{\citenamefont {Connes}\ and\ \citenamefont
  {Berberian}(1994)}]{Connes1994}%
  \BibitemOpen
  \bibfield  {author} {\bibinfo {author} {\bibfnamefont {A.}~\bibnamefont
  {Connes}}\ and\ \bibinfo {author} {\bibfnamefont {S.}~\bibnamefont
  {Berberian}},\ }\href {https://books.google.de/books?id=ZnGUGTa99Q0C} {\emph
  {\bibinfo {title} {Noncommutative Geometry}}}\ (\bibinfo  {publisher}
  {Elsevier Science},\ \bibinfo {year} {1994})\BibitemShut {NoStop}%
\bibitem [{\citenamefont {Bellissard}\ \emph {et~al.}(1994)\citenamefont
  {Bellissard}, \citenamefont {van Elst},\ and\ \citenamefont
  {Schulz-Baldes}}]{Bellissard1994}%
  \BibitemOpen
  \bibfield  {author} {\bibinfo {author} {\bibfnamefont {J.}~\bibnamefont
  {Bellissard}}, \bibinfo {author} {\bibfnamefont {A.}~\bibnamefont {van
  Elst}},\ and\ \bibinfo {author} {\bibfnamefont {H.}~\bibnamefont
  {Schulz-Baldes}},\ }\href@noop {} {\bibfield  {journal} {\bibinfo  {journal}
  {J. Math. Phys.}\ }\textbf {\bibinfo {volume} {35}},\ \bibinfo {pages} {5373}
  (\bibinfo {year} {1994})}\BibitemShut {NoStop}%
\bibitem [{\citenamefont {Moyal}(1949)}]{Moyal1949}%
  \BibitemOpen
  \bibfield  {author} {\bibinfo {author} {\bibfnamefont {J.~E.}\ \bibnamefont
  {Moyal}},\ }in\ \href@noop {} {\emph {\bibinfo {booktitle} {Math. Proc. Camb.
  Philos. Soc}}},\ Vol.~\bibinfo {volume} {45}\ (\bibinfo {organization}
  {Cambridge University Press},\ \bibinfo {year} {1949})\ pp.\ \bibinfo {pages}
  {99--124}\BibitemShut {NoStop}%
\bibitem [{\citenamefont {Groenewold}(1946)}]{Groenewold1946}%
  \BibitemOpen
  \bibfield  {author} {\bibinfo {author} {\bibfnamefont {H.~J.}\ \bibnamefont
  {Groenewold}},\ }in\ \href@noop {} {\emph {\bibinfo {booktitle} {On the
  Principles of Elementary Quantum Mechanics}}}\ (\bibinfo  {publisher}
  {Springer},\ \bibinfo {year} {1946})\ pp.\ \bibinfo {pages}
  {1--56}\BibitemShut {NoStop}%
\bibitem [{\citenamefont {{Khalkhali}}(2004)}]{Khalkhali2004}%
  \BibitemOpen
  \bibfield  {author} {\bibinfo {author} {\bibfnamefont {M.}~\bibnamefont
  {{Khalkhali}}},\ }\href@noop {} {\bibfield  {journal} {\bibinfo  {journal}
  {arXiv Mathematics e-prints}\ ,\ \bibinfo {eid} {math/0408416}} (\bibinfo
  {year} {2004})},\ \Eprint {https://arxiv.org/abs/math/0408416}
  {arXiv:math/0408416 [math.KT]} \BibitemShut {NoStop}%
\bibitem [{\citenamefont {Thouless}\ \emph {et~al.}(1982)\citenamefont
  {Thouless}, \citenamefont {Kohmoto}, \citenamefont {Nightingale},\ and\
  \citenamefont {den Nijs}}]{Thouless1982}%
  \BibitemOpen
  \bibfield  {author} {\bibinfo {author} {\bibfnamefont {D.~J.}\ \bibnamefont
  {Thouless}}, \bibinfo {author} {\bibfnamefont {M.}~\bibnamefont {Kohmoto}},
  \bibinfo {author} {\bibfnamefont {M.~P.}\ \bibnamefont {Nightingale}},\ and\
  \bibinfo {author} {\bibfnamefont {M.}~\bibnamefont {den Nijs}},\ }\href
  {https://doi.org/10.1103/PhysRevLett.49.405} {\bibfield  {journal} {\bibinfo
  {journal} {Phys. Rev. Lett.}\ }\textbf {\bibinfo {volume} {49}},\ \bibinfo
  {pages} {405} (\bibinfo {year} {1982})}\BibitemShut {NoStop}%
\bibitem [{\citenamefont {Qi}\ and\ \citenamefont {Zhang}(2011)}]{Qi2011}%
  \BibitemOpen
  \bibfield  {author} {\bibinfo {author} {\bibfnamefont {X.-L.}\ \bibnamefont
  {Qi}}\ and\ \bibinfo {author} {\bibfnamefont {S.-C.}\ \bibnamefont {Zhang}},\
  }\href@noop {} {\bibfield  {journal} {\bibinfo  {journal} {Rev. Mod. Phys.}\
  }\textbf {\bibinfo {volume} {83}},\ \bibinfo {pages} {1057} (\bibinfo {year}
  {2011})}\BibitemShut {NoStop}%
\bibitem [{\citenamefont {Qi}\ \emph {et~al.}(2008)\citenamefont {Qi},
  \citenamefont {Hughes},\ and\ \citenamefont {Zhang}}]{Qi2008}%
  \BibitemOpen
  \bibfield  {author} {\bibinfo {author} {\bibfnamefont {X.-L.}\ \bibnamefont
  {Qi}}, \bibinfo {author} {\bibfnamefont {T.~L.}\ \bibnamefont {Hughes}},\
  and\ \bibinfo {author} {\bibfnamefont {S.-C.}\ \bibnamefont {Zhang}},\
  }\href@noop {} {\bibfield  {journal} {\bibinfo  {journal} {Phys. Rev. B}\
  }\textbf {\bibinfo {volume} {78}},\ \bibinfo {pages} {195424} (\bibinfo
  {year} {2008})}\BibitemShut {NoStop}%
\bibitem [{\citenamefont {Sheikh-Jabbari}(2001)}]{Sheikh-Jabbari2001}%
  \BibitemOpen
  \bibfield  {author} {\bibinfo {author} {\bibfnamefont {M.}~\bibnamefont
  {Sheikh-Jabbari}},\ }\href@noop {} {\bibfield  {journal} {\bibinfo  {journal}
  {Phys. Lett. B}\ }\textbf {\bibinfo {volume} {510}},\ \bibinfo {pages} {247}
  (\bibinfo {year} {2001})}\BibitemShut {NoStop}%
\bibitem [{\citenamefont {Streda}(1982)}]{Streda1982}%
  \BibitemOpen
  \bibfield  {author} {\bibinfo {author} {\bibfnamefont {P.}~\bibnamefont
  {Streda}},\ }\href@noop {} {\bibfield  {journal} {\bibinfo  {journal} {J.
  Phys. C: Solid State Phys.}\ }\textbf {\bibinfo {volume} {15}},\ \bibinfo
  {pages} {L717} (\bibinfo {year} {1982})}\BibitemShut {NoStop}%
\bibitem [{\citenamefont {Xiao}\ \emph {et~al.}(2005)\citenamefont {Xiao},
  \citenamefont {Shi},\ and\ \citenamefont {Niu}}]{Xiao2005}%
  \BibitemOpen
  \bibfield  {author} {\bibinfo {author} {\bibfnamefont {D.}~\bibnamefont
  {Xiao}}, \bibinfo {author} {\bibfnamefont {J.}~\bibnamefont {Shi}},\ and\
  \bibinfo {author} {\bibfnamefont {Q.}~\bibnamefont {Niu}},\ }\href@noop {}
  {\bibfield  {journal} {\bibinfo  {journal} {Phys. Rev. Lett.}\ }\textbf
  {\bibinfo {volume} {95}},\ \bibinfo {pages} {137204} (\bibinfo {year}
  {2005})}\BibitemShut {NoStop}%
\bibitem [{\citenamefont {Xiao}\ \emph {et~al.}(2010)\citenamefont {Xiao},
  \citenamefont {Chang},\ and\ \citenamefont {Niu}}]{Xiao2010}%
  \BibitemOpen
  \bibfield  {author} {\bibinfo {author} {\bibfnamefont {D.}~\bibnamefont
  {Xiao}}, \bibinfo {author} {\bibfnamefont {M.-C.}\ \bibnamefont {Chang}},\
  and\ \bibinfo {author} {\bibfnamefont {Q.}~\bibnamefont {Niu}},\ }\href@noop
  {} {\bibfield  {journal} {\bibinfo  {journal} {Rev. Mod. Phys.}\ }\textbf
  {\bibinfo {volume} {82}},\ \bibinfo {pages} {1959} (\bibinfo {year}
  {2010})}\BibitemShut {NoStop}%
\bibitem [{\citenamefont {Banyai}\ and\ \citenamefont
  {Elsayed}(1994)}]{Banyai1994}%
  \BibitemOpen
  \bibfield  {author} {\bibinfo {author} {\bibfnamefont {L.}~\bibnamefont
  {Banyai}}\ and\ \bibinfo {author} {\bibfnamefont {K.}~\bibnamefont
  {Elsayed}},\ }\href@noop {} {\bibfield  {journal} {\bibinfo  {journal} {Ann.
  Phys.}\ }\textbf {\bibinfo {volume} {233}},\ \bibinfo {pages} {165} (\bibinfo
  {year} {1994})}\BibitemShut {NoStop}%
\bibitem [{\citenamefont {M{\"u}ller}\ \emph {et~al.}(2019)\citenamefont
  {M{\"u}ller}, \citenamefont {Hoffmann}, \citenamefont {Di{\ss}elkamp},
  \citenamefont {Sch{\"u}rhoff}, \citenamefont {Mavros}, \citenamefont
  {Sallermann}, \citenamefont {Kiselev}, \citenamefont {J{\'o}nsson},\ and\
  \citenamefont {Bl{\"u}gel}}]{Mueller2019}%
  \BibitemOpen
  \bibfield  {author} {\bibinfo {author} {\bibfnamefont {G.~P.}\ \bibnamefont
  {M{\"u}ller}}, \bibinfo {author} {\bibfnamefont {M.}~\bibnamefont
  {Hoffmann}}, \bibinfo {author} {\bibfnamefont {C.}~\bibnamefont
  {Di{\ss}elkamp}}, \bibinfo {author} {\bibfnamefont {D.}~\bibnamefont
  {Sch{\"u}rhoff}}, \bibinfo {author} {\bibfnamefont {S.}~\bibnamefont
  {Mavros}}, \bibinfo {author} {\bibfnamefont {M.}~\bibnamefont {Sallermann}},
  \bibinfo {author} {\bibfnamefont {N.~S.}\ \bibnamefont {Kiselev}}, \bibinfo
  {author} {\bibfnamefont {H.}~\bibnamefont {J{\'o}nsson}},\ and\ \bibinfo
  {author} {\bibfnamefont {S.}~\bibnamefont {Bl{\"u}gel}},\ }\href@noop {}
  {\bibfield  {journal} {\bibinfo  {journal} {Phys. Rev. B}\ }\textbf {\bibinfo
  {volume} {99}},\ \bibinfo {pages} {224414} (\bibinfo {year}
  {2019})}\BibitemShut {NoStop}%
\bibitem [{\citenamefont {Lux}\ \emph {et~al.}(2019)\citenamefont {Lux},
  \citenamefont {Freimuth}, \citenamefont {Bl{\"u}gel},\ and\ \citenamefont
  {Mokrousov}}]{Lux2019}%
  \BibitemOpen
  \bibfield  {author} {\bibinfo {author} {\bibfnamefont {F.~R.}\ \bibnamefont
  {Lux}}, \bibinfo {author} {\bibfnamefont {F.}~\bibnamefont {Freimuth}},
  \bibinfo {author} {\bibfnamefont {S.}~\bibnamefont {Bl{\"u}gel}},\ and\
  \bibinfo {author} {\bibfnamefont {Y.}~\bibnamefont {Mokrousov}},\ }\href@noop
  {} {\bibfield  {journal} {\bibinfo  {journal} {In preparation}\ } (\bibinfo
  {year} {2019})}\BibitemShut {NoStop}%
\bibitem [{\citenamefont {Meynell}\ \emph {et~al.}(2014)\citenamefont
  {Meynell}, \citenamefont {Wilson}, \citenamefont {Loudon}, \citenamefont
  {Spitzig}, \citenamefont {Rybakov}, \citenamefont {Johnson},\ and\
  \citenamefont {Monchesky}}]{Meynell2014}%
  \BibitemOpen
  \bibfield  {author} {\bibinfo {author} {\bibfnamefont {S.}~\bibnamefont
  {Meynell}}, \bibinfo {author} {\bibfnamefont {M.}~\bibnamefont {Wilson}},
  \bibinfo {author} {\bibfnamefont {J.}~\bibnamefont {Loudon}}, \bibinfo
  {author} {\bibfnamefont {A.}~\bibnamefont {Spitzig}}, \bibinfo {author}
  {\bibfnamefont {F.}~\bibnamefont {Rybakov}}, \bibinfo {author} {\bibfnamefont
  {M.}~\bibnamefont {Johnson}},\ and\ \bibinfo {author} {\bibfnamefont
  {T.}~\bibnamefont {Monchesky}},\ }\href@noop {} {\bibfield  {journal}
  {\bibinfo  {journal} {Phys. Rev. B}\ }\textbf {\bibinfo {volume} {90}},\
  \bibinfo {pages} {224419} (\bibinfo {year} {2014})}\BibitemShut {NoStop}%
\bibitem [{\citenamefont {K{\"u}bler}\ and\ \citenamefont
  {Felser}(2014)}]{Kuebler2014}%
  \BibitemOpen
  \bibfield  {author} {\bibinfo {author} {\bibfnamefont {J.}~\bibnamefont
  {K{\"u}bler}}\ and\ \bibinfo {author} {\bibfnamefont {C.}~\bibnamefont
  {Felser}},\ }\href@noop {} {\bibfield  {journal} {\bibinfo  {journal} {EPL
  (Europhysics Letters)}\ }\textbf {\bibinfo {volume} {108}},\ \bibinfo {pages}
  {67001} (\bibinfo {year} {2014})}\BibitemShut {NoStop}%
\bibitem [{\citenamefont {Nayak}\ \emph {et~al.}(2016)\citenamefont {Nayak},
  \citenamefont {Fischer}, \citenamefont {Sun}, \citenamefont {Yan},
  \citenamefont {Karel}, \citenamefont {Komarek}, \citenamefont {Shekhar},
  \citenamefont {Kumar}, \citenamefont {Schnelle}, \citenamefont {K{\"u}bler}
  \emph {et~al.}}]{Nayak2016}%
  \BibitemOpen
  \bibfield  {author} {\bibinfo {author} {\bibfnamefont {A.~K.}\ \bibnamefont
  {Nayak}}, \bibinfo {author} {\bibfnamefont {J.~E.}\ \bibnamefont {Fischer}},
  \bibinfo {author} {\bibfnamefont {Y.}~\bibnamefont {Sun}}, \bibinfo {author}
  {\bibfnamefont {B.}~\bibnamefont {Yan}}, \bibinfo {author} {\bibfnamefont
  {J.}~\bibnamefont {Karel}}, \bibinfo {author} {\bibfnamefont {A.~C.}\
  \bibnamefont {Komarek}}, \bibinfo {author} {\bibfnamefont {C.}~\bibnamefont
  {Shekhar}}, \bibinfo {author} {\bibfnamefont {N.}~\bibnamefont {Kumar}},
  \bibinfo {author} {\bibfnamefont {W.}~\bibnamefont {Schnelle}}, \bibinfo
  {author} {\bibfnamefont {J.}~\bibnamefont {K{\"u}bler}}, \emph {et~al.},\
  }\href@noop {} {\bibfield  {journal} {\bibinfo  {journal} {Sci. Adv.}\
  }\textbf {\bibinfo {volume} {2}},\ \bibinfo {pages} {e1501870} (\bibinfo
  {year} {2016})}\BibitemShut {NoStop}%
\bibitem [{\citenamefont {Seiberg}\ and\ \citenamefont
  {Witten}(1999)}]{Seiberg1999}%
  \BibitemOpen
  \bibfield  {author} {\bibinfo {author} {\bibfnamefont {N.}~\bibnamefont
  {Seiberg}}\ and\ \bibinfo {author} {\bibfnamefont {E.}~\bibnamefont
  {Witten}},\ }\href@noop {} {\bibfield  {journal} {\bibinfo  {journal} {J.
  High Energy Phys.}\ }\textbf {\bibinfo {volume} {1999}}\bibinfo  {number} {
  (09)},\ \bibinfo {pages} {032}}\BibitemShut {NoStop}%
\bibitem [{\citenamefont {Douglas}\ and\ \citenamefont
  {Nekrasov}(2001)}]{Douglas2001}%
  \BibitemOpen
\bibfield  {number} {  }\bibfield  {author} {\bibinfo {author} {\bibfnamefont
  {M.~R.}\ \bibnamefont {Douglas}}\ and\ \bibinfo {author} {\bibfnamefont
  {N.~A.}\ \bibnamefont {Nekrasov}},\ }\href@noop {} {\bibfield  {journal}
  {\bibinfo  {journal} {Rev. Mod. Phys.}\ }\textbf {\bibinfo {volume} {73}},\
  \bibinfo {pages} {977} (\bibinfo {year} {2001})}\BibitemShut {NoStop}%
\bibitem [{\citenamefont {Szabo}(2003)}]{Szabo2003}%
  \BibitemOpen
  \bibfield  {author} {\bibinfo {author} {\bibfnamefont {R.~J.}\ \bibnamefont
  {Szabo}},\ }\href@noop {} {\bibfield  {journal} {\bibinfo  {journal} {Phys.
  Rep.}\ }\textbf {\bibinfo {volume} {378}},\ \bibinfo {pages} {207} (\bibinfo
  {year} {2003})}\BibitemShut {NoStop}%
\end{thebibliography}
\end{document}